# THE POST-MERGER ELLIPTICAL NGC 1700: STELLAR KINEMATIC FIELDS TO FOUR EFFECTIVE RADII[1]

Thomas S. Statler[2,3], Tammy Smecker-Hane[4,5], and Gerald N. Cecil[2]

[2]Department of Physics and Astronomy, CB# 3255, University of North Carolina, Chapel Hill, NC 25799-3255

[3]Present address: Department of Physics and Astronomy, Clippinger Research Labs 251B, Ohio University, Athens, OH 45701

[4]Dominion Astrophysical Observatory, Herzberg Institute of Astrophysics, 5071 W. Saanich Rd., R. R. 5, Victoria, BC V8X 4M6, Canada

[5]Present address: Department of Physics, University of California, Irvine, CA 92717

## ABSTRACT

We have measured the stellar motions in the elliptical galaxy NGC 1700 along four position angles, to very large radii, using absorption features in spectra obtained with the Multiple Mirror Telescope. Our data extend the coverage of the stellar velocity field by a factor of 5 (2.5 times further in radius and twice as many PAs) beyond previous work. We have attained $10\,\mathrm{km\,s^{-1}}$ accuracy in the mean velocity $V$ out to nearly 2 effective radii ($r_e$), and errors are $< 15\%$ of the maximum rotation speed out to nearly $3\,r_e$. The lack of detectable minor-axis rotation and the nearly identical kinematics on the $\pm 45°$ PAs suggest that NGC 1700 is nearly oblate for $r \lesssim 2.5\,r_e$. Beyond this radius, twisting of the morphological and kinematic axes indicate increasing triaxiality, an intrinsic twist, or both. The velocity distribution in the low-amplitude counterrotating core is weakly skewed in the direction of rotation, arguing against a central stellar disk. The small skewness and the depression of the central velocity dispersion are consistent with the accretion of a low-mass stellar companion in a retrograde orbit. Photometric fine structure at large radii (Schweizer & Seitzer 1992) is also indicative of a merger; a velocity reversal $\sim 50''$ northeast suggests a major event. However, radially increasing prograde rotation in the main body of the galaxy implies that this was not the same event responsible for the counterrotating core. The strong rotation at large $R$ and the nearly oblate shape are consistent with $N$-body simulations of group mergers (Weil 1995); that all disturbances inside $\sim 2.5\,r_e$ have phase-mixed out suggests that NGC 1700 owes its present form to a merger of 3 or more stellar systems 2 – $4h^{-1}$ Gyr ago.

## 1. Introduction

The stellar motions in elliptical galaxies hold important clues to their origin and evolution. Specific physical processes such as major mergers (Toomre & Toomre 1972, Schweizer 1990, Barnes 1992, Hernquist 1993, Heyl *et al.* 1994) and accretion of small satellites (Kormendy 1984, Balcells & Quinn 1990, Balcells 1991) are expected to leave telltale kinematic signatures. More broadly, stellar kinematics are an indispensable diagnostic for intrinsic shapes (Contopoulos 1956, Binney 1985, Franx *et al.* 1991), which offer potential measures of the dynamical temperature of collapsing protogalaxies at turnaround (Aguilar & Merritt 1990), the amount of dissipation during collapse (Dubinski 1994), or even the number of systems involved in recent mergers (Weil 1995).

Recently, we have shown how tight constraints on the intrinsic shapes of individual ellipticals may be obtained by multiple-position-angle sampling of the mean stellar velocity field (VF), combined with surface photometry and dynamical model fitting (Statler 1994$a,b$). Though rotation is not the main source of support for luminous ellipticals, the VF bears the imprint of the critical points associated with orbit family boundaries, whose locations reveal the shape of the gravitational potential. Other measurable kinematic quantities are of course equally important. Primary support against gravity is provided by the velocity dispersion, radial profiles of which are crucial for virial mass estimates. Higher moments of the line-of-sight velocity distribution (LOSVD) gauge velocity anisotropy (Gerhard 1991), and their spatial variation has strong bearing on evidence for both central black holes (Merritt 1993) and dark halos (Merrifield & Kent 1990). In each case, kinematic data of very high quality are needed if fundamental structural parameters are to be obtained to better than factor of two, or even order of magnitude, accuracy.

Our previous work (Statler 1994$a$) showed that good shape constraints can be derived for giant ellipticals whose VFs are measured (1) on at least 4 position angles (PAs), (2) with an accuracy of $\sim 15\%$ of the maximum rotation speed (*i.e.,* $\sim 10\,\mathrm{km\,s^{-1}}$), and (3) outside the turnover radius of the rotation curve, which generally lies near the effective radius $r_e$. Unfortunately, there are no observations in the literature which fulfill all three requirements; deep long-slit observations are usually limited to one, or at most two, position angles (e.g. Saglia *et al.* 1993, Carollo *et al.* 1995), and multi-position-angle data tend not to reach to large radii (e.g. Davies & Birkinshaw 1988). In order to demonstrate that VFs can, and should, be mapped with better than $10\,\mathrm{km\,s^{-1}}$ accuracy beyond $r_e$, we undertook a pilot program to map the stellar kinematic fields of the E3 galaxy NGC 1700.

Like most elliptical galaxies, NGC 1700 is a superficially ordinary object with a few unique peculiarities. It is a fairly luminous giant elliptical, with absolute magnitude $M_B = -22.3$ (for $H_0 = 50\,\mathrm{km\,s^{-1}\,Mpc^{-1}}$), and central velocity dispersion $\sigma_0 \approx 230\,\mathrm{km\,s^{-1}}$ (Bender *et al.* 1992). Its rather high surface brightness, $\mu_B(r_e) \approx 21.2$ mag arcsec$^{-2}$, and uncharacteristically small effective radius, $r_e = 14'' = 2.6h^{-1}$ kpc (Franx *et al.* 1989*a*, Goudefrooij *et al.* 1994*a*), put it slightly less than $2\sigma$ off the Fundamental Plane (FP), in the sense of having unusually low $M/L$ for galaxies of this mass. Projected into the FP, it falls near the edge of the populated region by virtue of its compactness. Searches for interstellar matter have yielded only upper limits: $M(\mathrm{H\,I})/L_B < 0.021$ (Huchtmeier 1994), $M(\mathrm{H\,II}) < 1.7 \times 10^3\,\mathrm{M_\odot}$, no visible dust features and no detection in the IRAS 60 or 100$\mu$m bands (Goudefrooij *et al.* 1994*b*). Photometrically, NGC 1700 is unsurprising within $\sim 3\,r_e$ of the center; there is no isophotal twist $> 2^\circ$ and only a modest outward ellipticity gradient. However, beyond about $40''$ there is an abrupt $15^\circ$ twist of the major axis, accompanied by large negative values of the $a_4$ parameter. This outer parallelogram-shaped structure is easily visible on the Palomar Sky Survey prints, and has been likened to an outer ring inclined to the main body of the galaxy (Franx *et al.* 1991*a*) or to tidal tails (Seitzer & Schweizer 1990), though the physical relevance of these visual impressions is far from clear. The outer isophotal distortions have earned NGC 1700 a "fine structure parameter" $\Sigma = 3.70$ and a heuristic merger age estimate of 5.5 – 8.3 Gyr (Schweizer & Seitzer 1992). The velocity and dispersion fields have been sampled on the major and minor axes by Franx *et al.* (1989*b*), to about $20''$. Minor axis rotation is consistent with zero, but the major axis shows a weakly counterrotating core (*cf.* Saha & Williams 1994, Bender *et al.* 1994). The importance of counterrotation as an indicator of a past merger or accretion event has been widely emphasized (Kormendy 1984, Jedrzejewski & Schechter 1988, Bender 1988, Franx & Illingworth 1988, Schweizer 1990, Bender & Surma 1992).

Our program was thus designed with four objectives in mind: (1) to map the mean velocity field with $10\,\mathrm{km\,s^{-1}}$ or better accuracy to at least $2\,r_e$, in order to derive the intrinsic shape of the main body of the galaxy; (2) to measure velocity dispersion profiles to as large radii as possible, to constrain the existence of a dark halo; (3) to map the kinematics in the disturbed outer regions of the galaxy, to look for dynamical signatures associated with the isophotal twist or the box/ring structure; and (4) to measure the higher moments of the LOSVD in the inner regions, to clarify the nature of the counterrotating core. We find that, while the kinematic structure of the main body of the galaxy is highly symmetric, indicative of a well phase-mixed, nearly oblate system, the inner and outer regions do show evidence of past merger or accretion events. The extent of the relaxed regions will allow us to constrain the time since the last major event, which the data suggest may have been a contemporaneous merger of several objects.

The strengths of the stellar absorption features also give us valuable constraints on the mix of ages and metal abundances in NGC 1700. Analysis of the H$\beta$, Mg b and other weak metal absorption lines in the spectra and the constraints they place on the radial gradients in the stellar population will be discussed in a subsequent paper.

Section 2 of this paper describes the observations, reductions, the derivation of velocity profiles, and the error analysis, including systematic effects. Section 3 presents the results, and section 4 discusses the implication of the observed kinematic fields for the intrinsic shape and formation history of NGC 1700. Section 5 concludes.

## 2. Data

### 2.1. Observations

The spectra were obtained with the Multiple Mirror Telescope and the Red Channel Spectrograph (Schmidt, Weymann & Foltz 1989) on 16 – 17 December 1993 UT. The slit used was $1''\!.0 \times 180''$, and the detector was the 1200×800 Loral CCD (15$\mu$m pixels, 1 pix = $0''\!.3$, read noise = 7 $e^-$). To decrease read noise, the CCD was read binned $1 \times 4$ pixels in the dispersion $\times$ spatial dimensions. The 1200 grooves/mm grating provided spectra from $\lambda\lambda$ 4760 – 5630 Å with a dispersion of 0.72 Å/pix and a resolution of 2.2 Å. This spectral region contains H$\beta$, Mg b and numerous weak metallic absorption lines.

Conditions were not photometric; the first night was complicated by varying weather ranging from heavy cloud to light cirrus. On the second night conditions varied from light cirrus to clear skies. Seeing also varied, but was in the vicinity of $1''\!.0$ – $1''\!.4$ for $\gtrsim 80\%$ of the time; some exposures were paused during brief periods of poor ($\gtrsim 3''$) seeing. Spectra of NGC 1700 were taken at four position angles: PA = 0 (minor axis), 225, 270 (major axis) and 315 degrees. Total exposure times on the four PAs were respectively 2.3, 2.0, 0.9, 2.0 hrs and are sums of either two (for PA 270) or four individual exposures. These times were chosen based on the intrinsic surface brightness along the axes and the atmospheric transparency at the time of the observations to produce a similar $S/N$ ratio on each PA. Spectra of separate blank sky positions were taken for sky subtraction because the galaxy was large enough to fill the slit. Three sky positions for PA = 0, 225/315 and 170 were chosen by inspection of the Palomar Sky Survey plates. A separate sky position for each PA was used in order to minimize sky subtraction errors that might otherwise have been introduced by flexure of the spectrograph. Two or three 900 s exposures at the sky position were interspersed among the NGC 1700 exposures at each PA. Spectra of He-Ar-Fe-Ne arc lamps were taken before or after each individual exposure.

Spectra of stars in the IAU radial velocity standards list and in M67 (Mathieu *et al.* 1986) were taken for radial velocity calibrations. In addition, spectra of stars with a range of gravity, temperature and metallicity were obtained in order to create composite templates to use in the cross correlation of the galaxy spectra. These stars were chosen from various sources and are either standards on which the Lick indices are based (Worthey *et al.* 1994) or have been measured on this system. In a separate paper, we will use these to investigate the age and metal abundance gradients in the stellar population of NGC 1700. All stars were trailed across the slit to illuminate it uniformly. Other observations used to calibrate the CCD consisted of standard bias frames, dark frames, quartz and twilight flats.

### 2.2. Reduction

#### *2.2.1. Initial Procedures*

All data were reduced using IRAF, following standard practice. After overscan and bias corrections, flat-fielding was done using combined dome and quartz lamp flats and an illumination correction derived from twilight sky spectra. Dark count was implicitly removed in sky subtraction. Cosmic rays were removed from the galaxy and sky frames by subtracting median images and filtering at a $5\sigma$ threshold. Wavelength calibration was performed with $\sim$ 28 lines from the He-Ne-Fe-Ar arc exposures. Spectra were straightened in the spatial direction using star traces at different positions along the slit. The galaxy frames were registered by fitting the central part of the spatial profile. One-dimensional stellar spectra were extracted using the standard procedures, with "unweighted extraction." All spectra were rebinned logarithmically in the dispersion direction to the same number of pixels, one pixel corresponding to $\Delta x \equiv \Delta \log \lambda = 6.764 \times 10^{-5}$, or $\Delta v = 46.689\,\mathrm{km\,s^{-1}}$.

### *2.2.2. Sky Subtraction*

Accurate sky subtraction was essential because our goal was to obtain velocity profiles at radii where the galaxy's surface brightness amounts to only 10 – 20% of sky. For each galaxy exposure, we formed a 2-dimensional image of the sky spectrum from a linear combination of two sky frames taken before and after, as follows. Let $Y(t_1)$ and $Y(t_2)$ be the two sky frames, taken at times $t_1$ and $t_2$, and $G(t)$ be the galaxy frame taken at time $t$; times refer to the middle of the exposures. The combined sky image was $Y_c = K[aY(t_1) + (1-a)Y(t_2)]$, where nominally $a = (t_2 - t)/(t_2 - t_1)$ and $K$ was the ratio of exposure times. In practice, however, we fine-tuned both $a$ and $K$ so as to obtain the best removal of the dominant sky emission lines. Fine tuning of $a$ and $K$ by $\sim$ 10% was sufficient to optimize the subtraction in most cases, though for exposures near the beginning or end of the night, when sky conditions were changing rapidly, 30% changes in the coefficients were sometimes necessary. By-eye optimization fixed the coefficients to within 2 – 5%; larger variations noticeably degraded the removal of the emission lines. Systematic errors introduced by sky subtraction were subsequently estimated by changing the coefficients and repeating the analysis (see section 2.3.5).

Averaging the sky spectra in the spatial direction was necessary to increase the signal-to-noise ratio of the sky determination. We therefore smoothed the spectrum spatially with a variable-width boxcar window, whose width increased from 1 pixel at the center of the galaxy to 15 pixels at the ends of the slit. After subtracting the smoothed composite sky image from each galaxy frame, the frames for each PA were co-added, producing four two-dimensional spectra.

### *2.2.3. Scattered Light*

Given the steep surface brightness profile of NGC 1700, we were concerned that a low-level background due to light scattered from the inner regions of the galaxy could produce spurious gradients in line strengths and velocity dispersion. We created a composite stellar profile to estimate the scattered light contribution. Normalized to unit flux within 60 pixels of the center, the composite profile can be fitted by a sum of a central Gaussian seeing disk of width $\sigma_{\mathrm{seeing}} = 1\,\mathrm{pixel} = 1''.2$ containing 79% of the light and a scattered component $f_{\mathrm{sc}}(y)$ given by

$$f_{\mathrm{sc}}(y) = \begin{cases} f_-(y), & y < -8, \\ \left[f_-(y) + f_+(y)\right]/2, & -8 \le y \le 8, \\ f_+(y), & y > 8, \end{cases} \tag{1}$$

where $y$ is distance from the center, measured in pixels, and

$$f_-(y) = 0.098(|y| + 2.34)^{-1.48}, \qquad f_+(y) = 0.146(|y| + 2.97)^{-1.61}. \tag{2}$$

The numerical constants in $f_-$ and $f_+$ are quite insensitive to $\sigma_{\rm seeing}$ or the assumed amplitude of the central Gaussian. We find that 10% of the light falls more than 5 pixels (6″) from the center; and since the galaxy's surface brightness profile is actually *steeper* than $f_{\rm sc}(y)$ at large $y$, scattered light from the bright center contributes an increasing fraction, of order 20%, of the local measured light beyond 30″.

Since our main concern was with a spurious continuum, we made a rough estimate of the light scattered from the galactic center, as follows. For each PA, the composite stellar profile was scaled to have the same total flux in the central 3 pixels as the galactic profile (averaged over wavelength). The galactic scattered light profile was then taken to be the identically scaled $f_{\rm sc}(y)$, with a wavelength dependence given by a low-order polynomial fit to the central galactic spectrum. The resulting scattered-light image was then subtracted from the sky-subtracted spectrum.

Repeating the LOSVD analysis produced results nearly identical to those from the spectra uncorrected for scattered light. We quantify the systematic effect in section 2.3.5; but in the end, the kinematic results were not significantly changed. The results in section 3 all include the correction described above.

## 2.3. LOSVD Extraction

### *2.3.1. Preliminaries*

Galaxy spectra were cut into spatial slices 1 pixel wide within 8″ of the center, and of increasing width at larger radii. The radial bin size was set by requiring $\gtrsim 1000$ counts per (wavelength) pixel in the binned spectra. Small intervals, 3 pixels wide, contaminated by stars on the slit at PAs 0 and 270 were removed. Data beyond 74″ could not be used because of serious vignetting. The outermost bin thus turned out narrower than those immediately interior, and so had correspondingly lower $S/N$. Regions contaminated by imperfectly subtracted night sky emission lines *e.g.,* $\lambda\lambda$ 5461 and 5577 Å, were excised and replaced with linear interpolations plus Gaussian noise. A parallel analysis where the entire spectrum longward of 5455 Å was discarded showed no noticeable difference.

Each spectrum was divided by a continuum fit with as high an order spline function as was consistent with the noise level; the IRAF 'order' parameter was 6 for quiet and 4 for noisy spectra. Changing the fitting function did not affect the results, since the normalized spectra were then filtered in the Fourier domain to remove low-frequency components. The filter was zero below a threshold wavenumber $k_L$, unity above $2k_L$, and joined by a cosine taper in between. Mean velocities and dispersions were found to be largely independent (to within $\lesssim 3\,{\rm km\,s^{-1}}$) of the threshold for $k_L$ in the range 0.011 – 0.019 pixel$^{-1}$; we adopted $k_L = 0.015$ pixel$^{-1}$, corresponding to a wavelength of 65 pixels in spectra padded out to a length of 1300 pixels.

### *2.3.2. Methods*

At any point along the slit, the galaxy spectrum $G(x)$ is the convolution of an ideal template $I(x)$, which represents the true stellar makeup of the galaxy at zero velocity, with the broadening function $B(x)$, which is just the normalized LOSVD written as a function of $x = v/c$. If one has an observed template $S(x)$ that is a good match to $I(x)$, then the galaxy-star cross-correlation function $X(\delta) = \int dx\, G(x+\delta)S(x)$ is the convolution of the template autocorrelation function $A(\delta) = \int dx\, S(x+\delta)S(x)$ with the broadening function.

This is the basis of the cross-correlation (XC) method (Simkin 1974, Tonry & Davis 1979). Here we used a modified form of the XC method which allows for non-Gaussian broadening functions. In this approach, the correlation function itself, rather than the observed galaxy spectrum, is fitted by the model line profile. As long as the galactic spectrum is well resolved (as is the case here), the results are robust and fairly insensitive to template mismatch, the formal errors accurately represent the uncertainties in the fitted parameters, and the standard reduced $\chi^2$ is a good measure of the goodness of fit of the model LOSVD profile, again largely decoupled from the fit of the spectral template. The method is fully described and tested by Statler (1995).

We adopted the parametrization introduced by van der Marel & Franx (1993, hereafter vdMF), in which the LOSVD is represented by a truncated Gauss-Hermite series:

$$L(v) = \frac{\gamma}{(2\pi)^{1/2}\sigma}\left[1 + h_3\frac{(2w^3-3w)}{3^{1/2}} + h_4\frac{(4w^4-12w^2+3)}{24^{1/2}}\right]e^{-w^2/2}, \qquad w \equiv \frac{v-V}{\sigma}. \tag{3}$$

Without the second and third terms in brackets, equation (3) is a Gaussian of dispersion $\sigma$ and amplitude $\gamma$ centered at the mean velocity $V$. The $h_3$ and $h_4$ terms describe deviations from Gaussian form that are antisymmetric and symmetric about the center, respectively, and are proportional to the coefficients of skewness and kurtosis. We made one change to vdMF's formulation, in that we forced $B(x) \equiv L(cx)$ to be nonnegative by truncating the oscillating tails of the distribution (which arise when $h_3$ or $h_4$ is nonzero) beyond the first zeros either side of the center. Since there is nothing special about the Gauss-Hermite parametrization beyond mathematical elegance, this added bit of physical realism comes at no cost. In the error analysis the error $\sigma_G$ per pixel in each normalized galactic spectrum was taken to be the RMS difference between the actual spectrum and a copy smoothed with a 7 pixel wide boxcar window. Tests with Gaussian noise added to artificially broadened stellar spectra verified that this algorithm could fairly faithfully recover the noise amplitude in the range of expected values. The same $\sigma_G$ applied to all pixels in the spectrum; a more careful pixel-by-pixel estimation of the noise produced results not noticeably different.

In section 3.2 we also present LOSVDs derived by the Fourier Correlation Quotient (FCQ) method of Bender (1990*a*). This method produces a nonparametric representation of the velocity distribution, but relies on filtering the high-frequency component of the data, and hence can be used only in regions of high $S/N$. We show below that the parametric and nonparametric extractions of the LOSVD agree with each other, so our results are not method-dependent.

### *2.3.3. Composite Templates*

The velocity zero point of our data was determined from 12 spectra of IAU radial velocity standards (*Astronomical Almanac*) and M67 stars of known velocity (Mathieu *et al.* 1986). These spectra were redshifted to zero heliocentric velocity; then velocities of the remaining stars were determined by cross-correlation with each of the IAU standards and M67 stars, and the unweighted mean was used to shift them to zero velocity.

We then carried through the LOSVD extraction for all of the galactic spectra using all possible templates. This was done twice, using both Gaussian and Gauss-Hermite broadening functions. We found the resulting profiles, at least out to moderate radii, to be highly symmetric (or antisymmetric) about the central values on each position angle (except PA 225, probably due to a positioning error; see section 2.3.4 below). Major-axis profiles of $V$ derived from the Gaussian fits and $h_3$ from the Gauss-Hermite fits are shown in figure 1. Using a different template results mostly in a constant offset of the kinematic parameters; this is consistent with the finding of van der Marel *et al.* (1994).

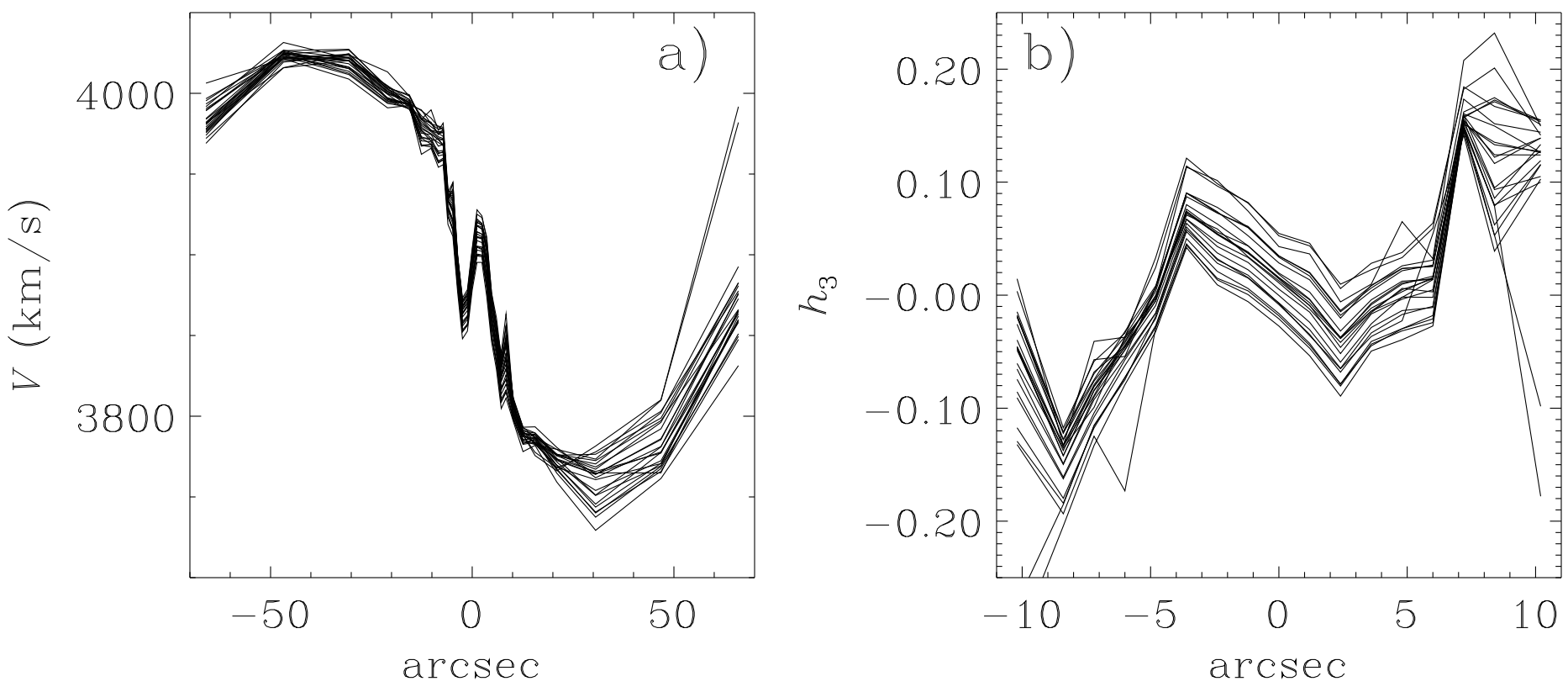

Fig. 1.— Major-axis profiles of two kinematic parameters, derived using all templates. (*a*) Mean velocity $V$ from Gaussian fits to the LOSVD. (*b*) The $h_3$ Gauss-Hermite coefficient, in the region where it is statistically significant. For the most part, changing the template results mainly in a constant offset. In (*b*), the profile using M67 S-488, an extreme giant, as template is omitted because it is very discrepant.

As shown by Rix & White (1992), systematic effects of template mismatch can be mitigated by constructing a composite template from a set of individual stellar spectra so that the broadened template reproduces the galactic spectrum as closely as possible. We found that our template spectra were sufficiently similar that the mathematical problem of trying to match the galaxy with a linear combination of any more than 4 of them was hopelessly degenerate. Appealing to the fact that the $h_3$ profile should be an odd function of radius along any cut through the center for an equilibrium system, we chose 4 stars with a range of spectral types from G dwarfs to K giants and metallicities near solar such that the central $h_3$ values obtained using each of them separately averaged near zero. The adopted stars were HD 117176 (G4V, [Fe/H] = −0.11), SAO 11983 (K0V, 0.36), HD 95345 (K1III, −0.05), and M67 S-1557 (gK3,[6] 0.0). Then, independently for each radial bin, we least-squares fit a linear combination of the normalized and filtered stellar spectra, shifted and broadened by the mean $V$ and $\sigma$ obtained from the individual Gaussian fits, to the normalized and filtered galactic spectrum. Even with as few as 4 stars, the optimization problem was sufficiently ill-conditioned that normally reliable algorithms such as Lucy's Method failed to converge in reasonable time. We therefore used a random search of the parameter space, which was 3-dimensional allowing for arbitrary normalization. Sampling $4.5 \times 10^4$ points in that space gave reproducible results for the composite templates. The adopted spectra for positions near the galactic center were made up of roughly 65% K3 giant, 26% G dwarf, 8% K dwarf, and 1% K1 giant. Outside $r_e$, the average proportions were 34% late K giant, 31% early K giant, 22% K dwarf, and 13% G dwarf; but there were substantial point-to-point fluctuations in the coefficients, reflecting the degeneracy of the spectral synthesis and the attendant sensitivity to observational noise.

Both the Gaussian and Gauss-Hermite LOSVD extractions were then repeated, using the appropriate composite template for each galactic spectrum, to yield the final kinematic profiles for the galaxy.

[6]Estimated from photometry of Janes & Smith (1984); MK class not available.

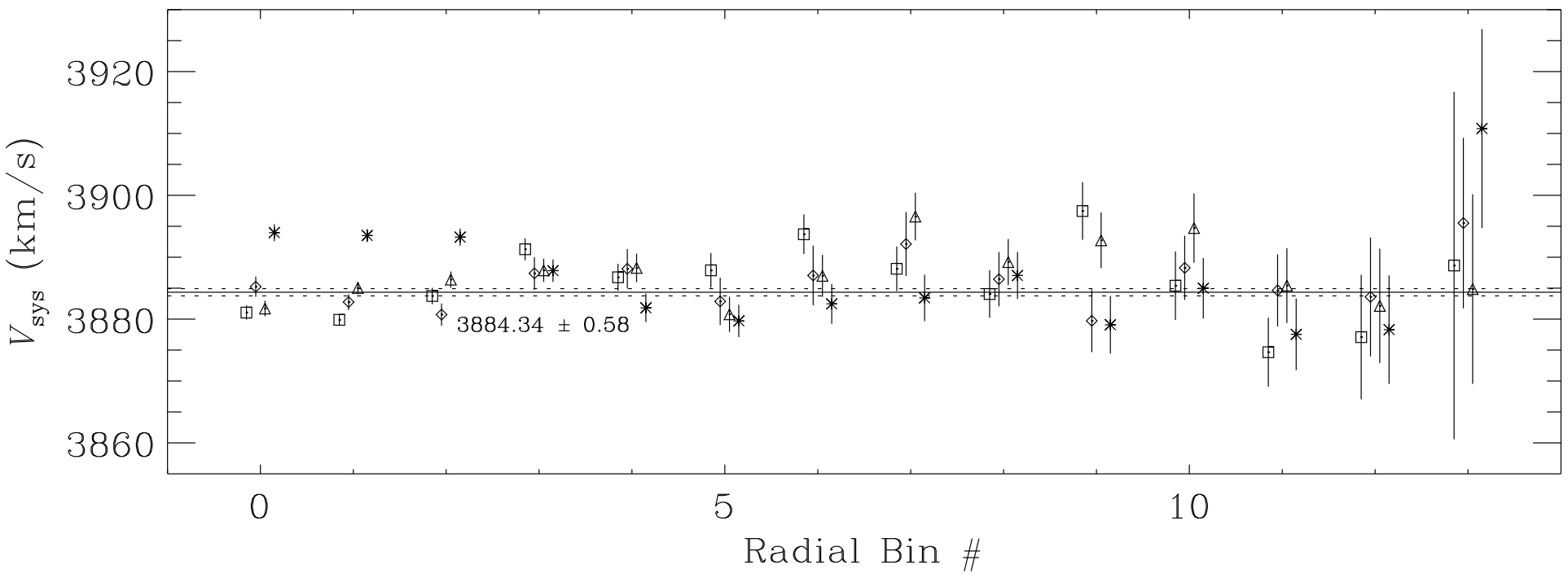

Fig. 2.— Estimates of the systemic velocity found by averaging velocities at positive and negative radii are plotted against radial position (as bin number) for all four position angles. *Squares*, PA 0; *diamonds*, PA 270, *triangles*, PA 315; *asterisks*, PA 225. Horizontal line indicates the weighted mean, omitting the PA 225 data (see text for full discussion).

### *2.3.4. Systemic Velocity*

At any projected distance $R$ from the center of the galaxy, $[V(R) + V(-R)]/2$ is an estimate of the systemic velocity. We computed this quantity for all radial bins except the outermost on all PAs. The results are shown in figure 2. That the data (excepting that for PA 225) are consistent with the same systemic velocity at all radii is an indication that the velocity field is, to good accuracy, point-symmetric across the center, as it should be in an equilibrium system.

On PA 225, the systemic velocity appears to increase at small $R$. One can see in figure 6 that the measured $V$ at the center is $\sim 15\,\mathrm{km\,s^{-1}}$ higher than the central velocity measured on the other PAs, which are all consistent with each other, and the PA 225 velocity profile is not antisymmetric about the central value. This is consistent with a mispositioning of the galaxy center by of order the slit width, and not surprising since PA 225 was observed at the beginning of night 2 through a large airmass. We therefore derived the systemic velocity by a weighted mean of the measurements shown in figure 2, omitting all the PA 225 data (though including PA 225 changes the result by only $1\,\mathrm{km\,s^{-1}}$); the result is $V_{\rm sys} = 3884.34 \pm 0.58\,\mathrm{km\,s^{-1}}$.

### *2.3.5. Systematic Errors*

The covariance matrix returned by the cross-correlation analysis gives a good estimate of the formal uncertainties, but does not include systematic errors. We attempted to make liberal estimates of the dominant systematic effects, which arise from template mismatch, sky subtraction, and scattered light, in decreasing order of importance.

Because the cross-correlation analysis was done for each individual and composite template, it was easy to measure the variance induced by template mismatch. Not all templates were included in this variance, however, since some were obviously poor representations of the galactic spectra, and were not used, or even considered, in the composites. Generally, bad templates were identified by the anomalously high dispersions that they produced; the variance was thus determined at each point from only those templates that produced $\sigma < 300\,\mathrm{km\,s^{-1}}$ on that PA. Template variances, normalized to the formal internal errors, are shown in figure

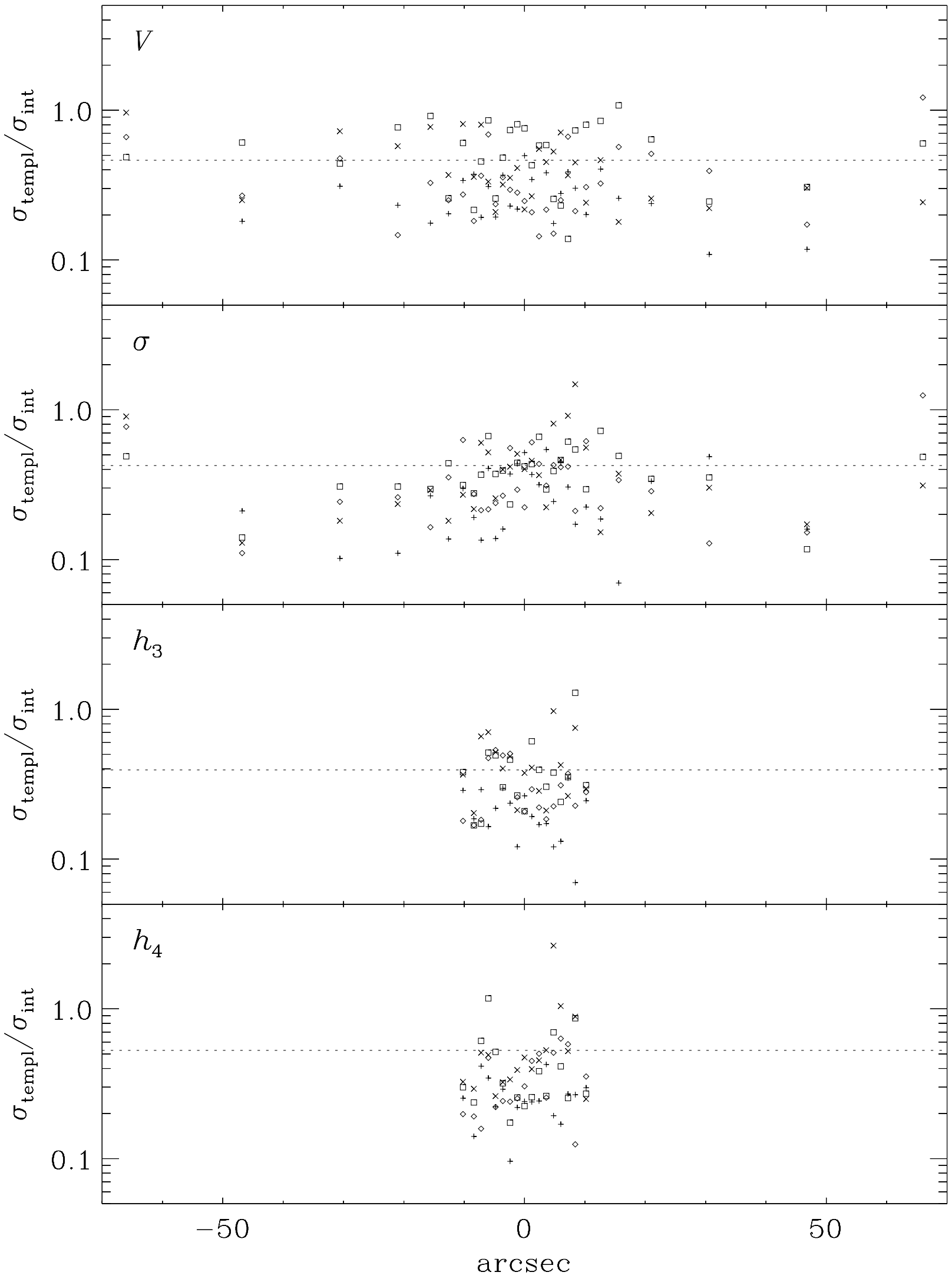

Fig. 3.— Variances, induced by the adopted template, in $V$, $\sigma$, $h_3$, and $h_4$, normalized to the corresponding internal errors, plotted against position along the slit. *Plus signs,* PA 0; *diamonds,* PA 225; *squares,* PA 270; *crosses,* PA 315.

3. Different symbols indicate the different slit PAs. Except in the velocity dispersion, there is no significant radial variation of the normalized variance. We formed an overall estimate of the systematic error in each parameter by adopting the RMS normalized variance over all measured points as a correction to the formal error to be applied below. These values are indicated by the dotted lines in the figure and are given in the first row of Table 1.

Gauging the error due to sky subtraction is more involved. We took PA 225 as being representative of the whole data set and reanalyzed only this part of the data. As before, sky frames were created for each galaxy exposure, but this time by the "maximally naive" method of choosing the sky exposure nearest in time, rather than by constructing a weighted average. The rest of the analysis proceeded exactly as before. The difference in the two methods, again normalized to the formal internal error, is shown in figure 4. We adopted the RMS normalized difference as the error correction factor for sky subtraction; the values are shown as the dotted lines in the figure and are given in the second row of Table 1.

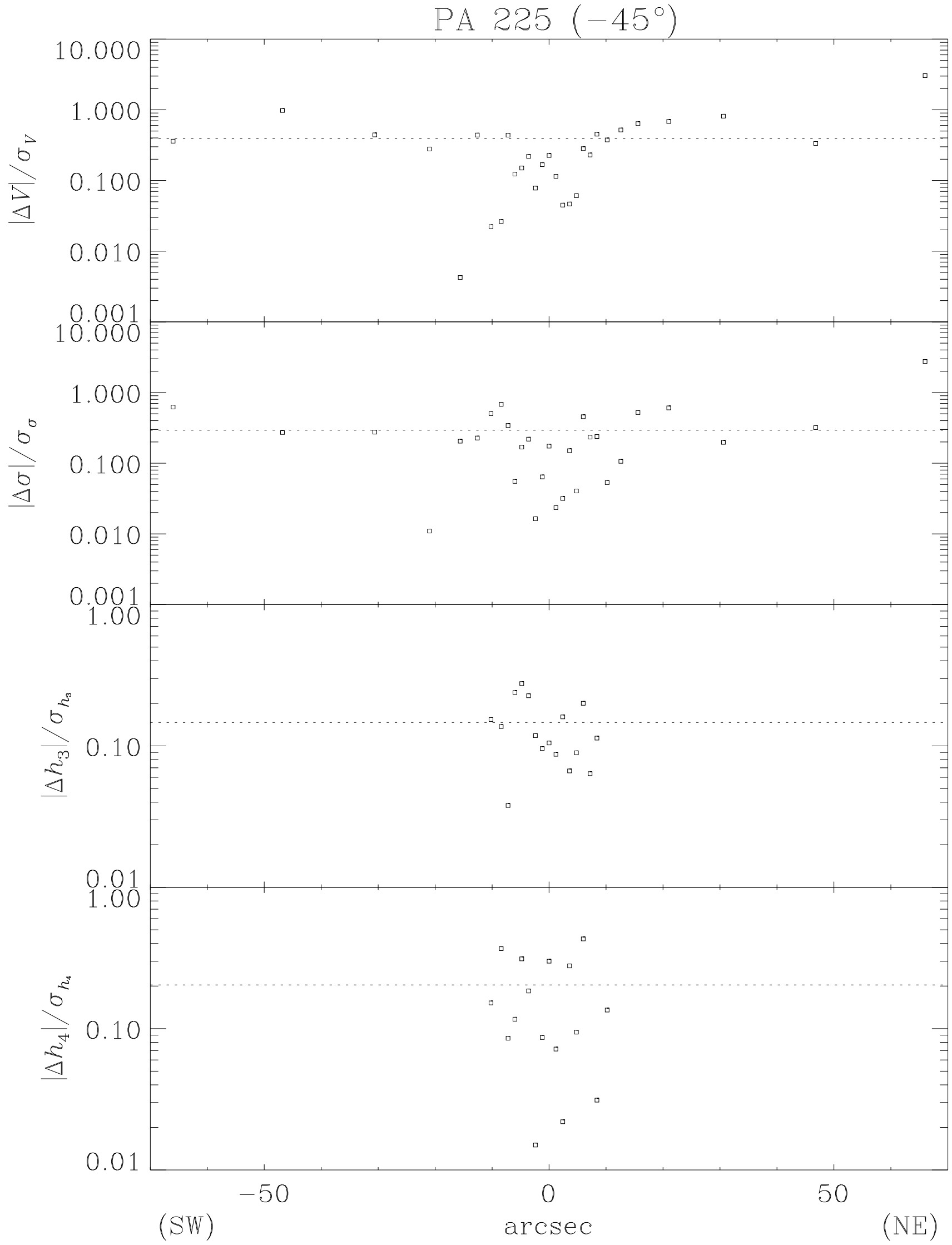

Fig. 4.— Difference in $V$, $\sigma$, $h_3$, and $h_4$, normalized to the internal errors, induced by the scattered light correction in the PA 225 data.

Similarly, the error due to scattered light was estimated as the RMS difference between the results obtained with and without the scattered light correction described in section 2.3.3. The values are listed in the third row of Table 1. Finally, the total error in each kinematic parameter was taken to be the quadrature sum of the internal error and the RMS systematic errors as estimated above. Thus, for $V$,

$$\sigma_{\rm total} = \sigma_{\rm int}(1^2 + .395^2 + .463^2 + .117^2)^{1/2} = 1.18\sigma_{\rm int}; \tag{4}$$

and so each error bar on $V$ was increased by a factor of 1.18 in all of the presentation that follows. The corresponding correction factors for the other kinematic parameters are given in the last row of Table 1. On the outermost data point at $+66''$ on PA 225, we further increased the error bars on $V$ and $\sigma$ by a factor of 3 to reflect the extreme sensitivity to sky subtraction apparent in figure 4. This outermost point suggests a velocity reversal, and will be discussed further in section 3.3.

## 3. Results

### 3.1. Parametrized Velocity Distributions

Profiles of the LOSVD parameters are shown in figures 5, 6, 7, and 8, and their values are listed in columns 2, 4, 6, and 8 of Table 2. Errors quoted in the Table are total errors, estimated as described above. Comparing the reduced $\chi^2$ of the Gaussian and Gauss-Hermite fits, we find that the latter are a statistically significant improvement over the former only within $11''$ of the center. The tabulated $h_3$ and $h_4$ values outside this region are italicized to indicate this, and the corresponding $V$ and $\sigma$ are obtained with Gaussian broadening functions. Note that the differences of $h_3$ and $h_4$ from zero in these outer regions are of marginal significance. In the figures, $h_3$ and $h_4$ are plotted only where the Gauss-Hermite fits are significant, for clarity.

There is no statistically significant rotation seen on the minor axis; inside $40''$, the maximum rotation speed is on the major axis and the velocity profiles on the diagonal position angles are nearly identical. In these respects, NGC 1700 resembles a "classic major axis rotator". It does, however, have a low-amplitude counterrotating core, previously noted by Franx *et al.* (1989*b*), Saha & Williams (1994), and Bender *et al.* (1994). The core's apparent rotation is also about the minor axis; it is interesting that there is no kinematic signature of the core at all detectable on the minor axis. At larger radii the velocity field becomes more complicated, but we will defer discussion of this to Sec. 3.3.

Table 1. RMS Systematic Errors (normalized to internal error)

| Source | $V$ | $\sigma$ | $h_3$ | $h_4$ |
|---|---|---|---|---|
| Template mismatch | 0.463 | 0.424 | 0.394 | 0.528 |
| Sky subtraction | 0.395 | 0.294 | 0.147 | 0.204 |
| Scattered light | 0.117 | 0.118 | 0.090 | 0.106 |
| $\frac{\text{Total error}}{\text{Internal Error}}$ | 1.18 | 1.13 | 1.09 | 1.15 |

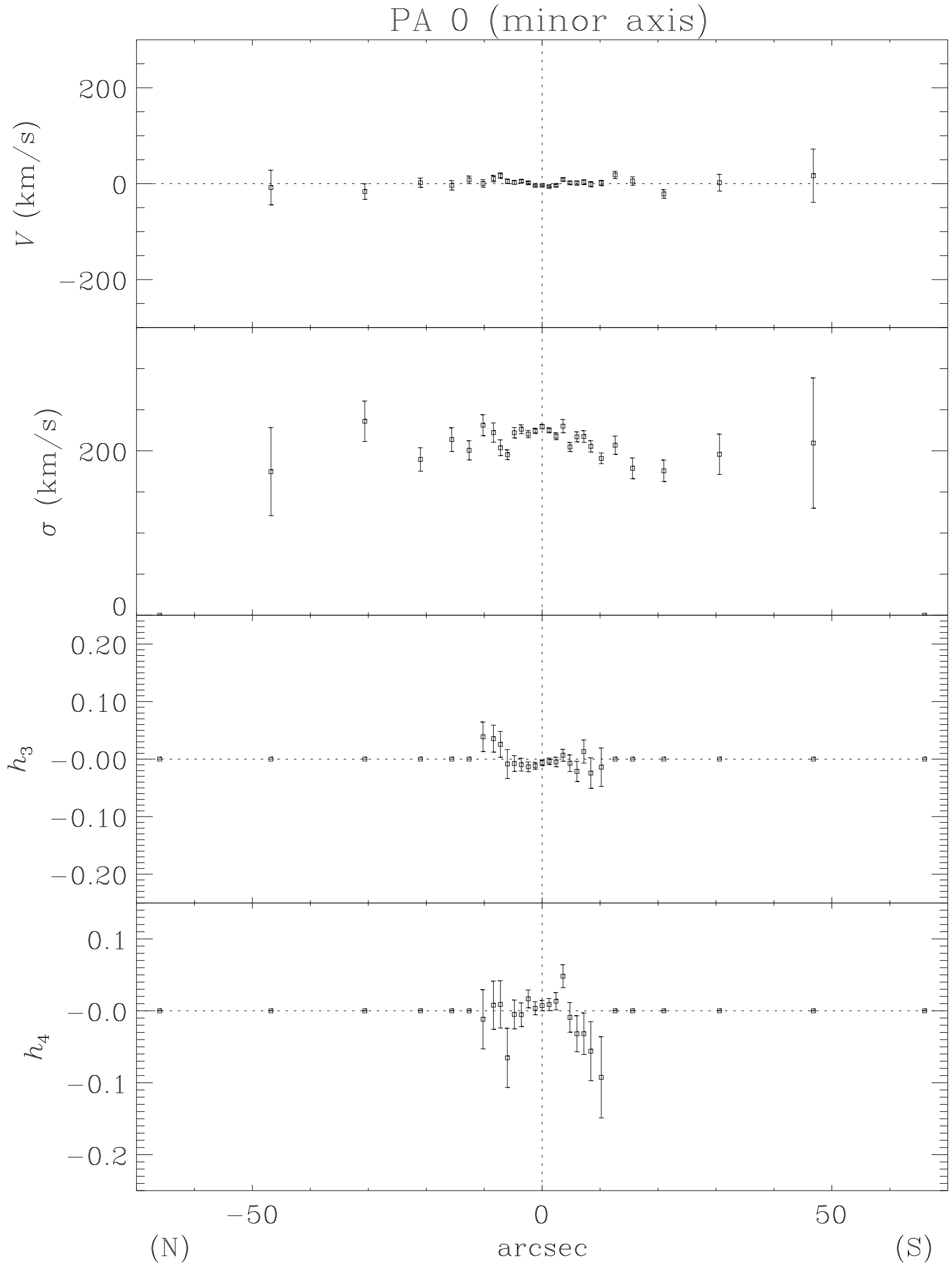

Fig. 5.— Kinematic profiles for the minor axis (PA 0) of NGC 1700. $V$, $\sigma$, $h_3$, and $h_4$ are the parameters in the Gauss-Hermite representation of the broadening function, eq. (3). Beyond $11''$, $h_3$ and $h_4$ are not statistically significant and the broadening function is assumed Gaussian. Note that there is no detectable rotation on the minor axis.

The major axis $h_3$ profile indicates that the LOSVD is slightly skewed in the direction of the mean rotation. This is consistent with other ellipticals for which non-Gaussian LOSVDs have been obtained (Bender 1990*a*, Rix & White 1992, van der Marel *et al.* 1994, Saha & Williams 1994). The sign of $h_3$ reverses with the reversal of the mean rotation in the core. The abruptness of the velocity reversal can also be seen in the $\sigma$ and $h_4$ profiles. The fitted $\sigma$ peaks at the edge of the core, consistent with superposition (both by projection and seeing) of direct and retrograde components. Negative values of $h_4$ indicate that the distribution is slightly platykurtic (*i.e.,* square-shouldered), as one would expect from such a superposition. All of these characteristics can be seen, albeit at lower contrast, on PAs 225 and 315 as well.

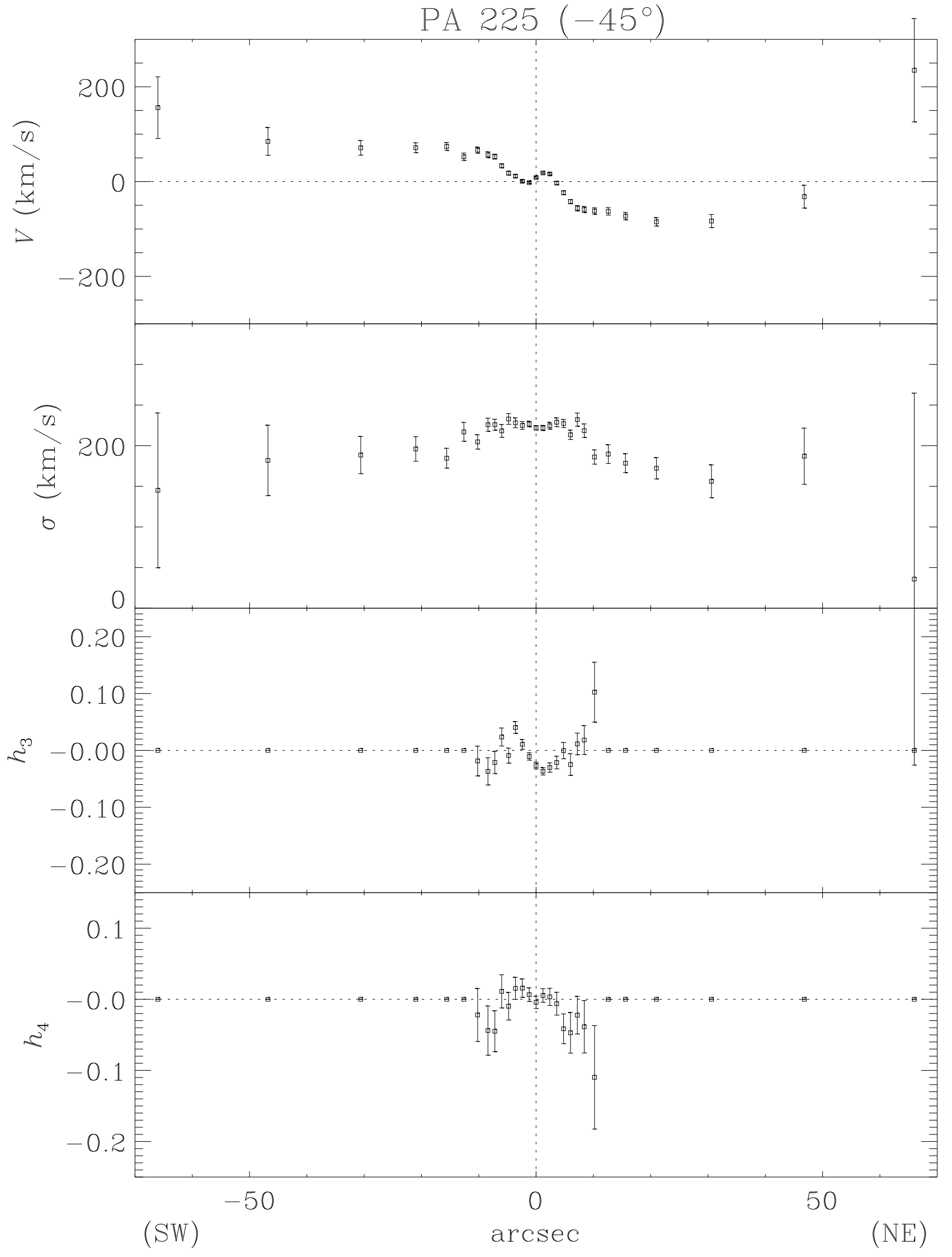

Fig. 6.— As in figure 5, but for one of the diagonal slit positions (PA 225). The lack of antisymmetry of the $V$ and $h_3$ profiles is likely due to mispositioning of the galactic center on the slit.

Reconstructed LOSVD profiles for the innermost 22″ of the major axis are plotted as the solid curves in in figure 9. Note the very modest skewness, especially in the counterrotating core. This contrasts sharply with the strong non-Gaussian distortions seen in the core of IC 1459 (Franx & Illingworth 1988); unlike that galaxy, there is not compelling evidence that the source of the counterrotation in NGC 1700 is a central

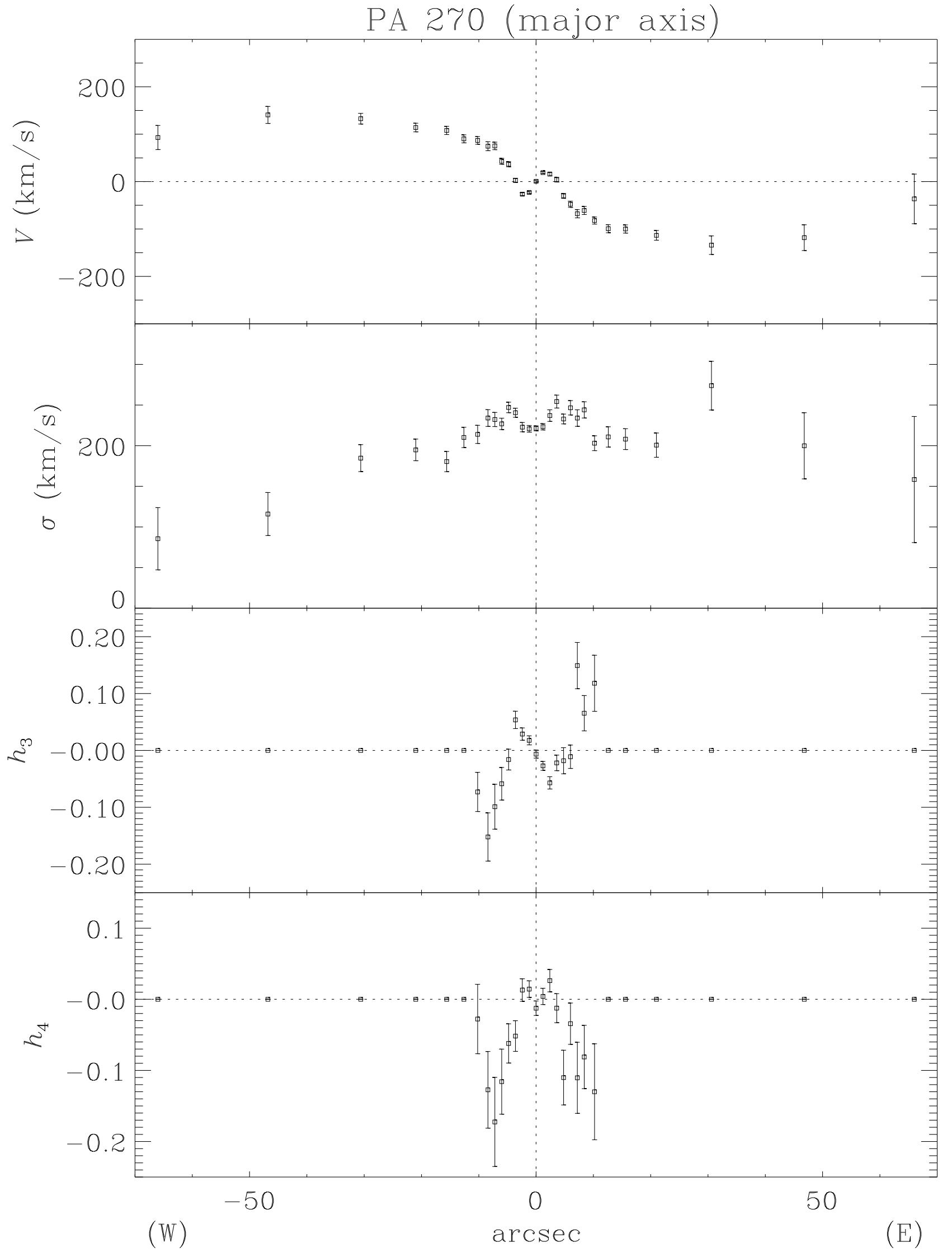

Fig. 7.— As in figure 5, but for the major axis (PA 270).

stellar disk.

### 3.2. Nonparametric Velocity Distributions

In the region where the $h_3$ and $h_4$ terms are statistically significant, we can compare the LOSVDs derived by parametric fits of the cross-correlation peaks to nonparametric extractions using the Fourier Correlation Quotient (FCQ) technique of Bender (1990*a*). This latter approach has the advantage of not relying on an assumed functional form for the LOSVD, but instead requires filtering of the high-frequency components from the data, and can therefore produce ambiguous results. Bender advocates the use of an optimal Wiener filter, whose transmission drops smoothly to zero at a wavenumber derived from extrapolating the power spectrum. In practice, however, we found it more effective to choose the cutoff wavenumber empirically to minimize Fourier "ringing."

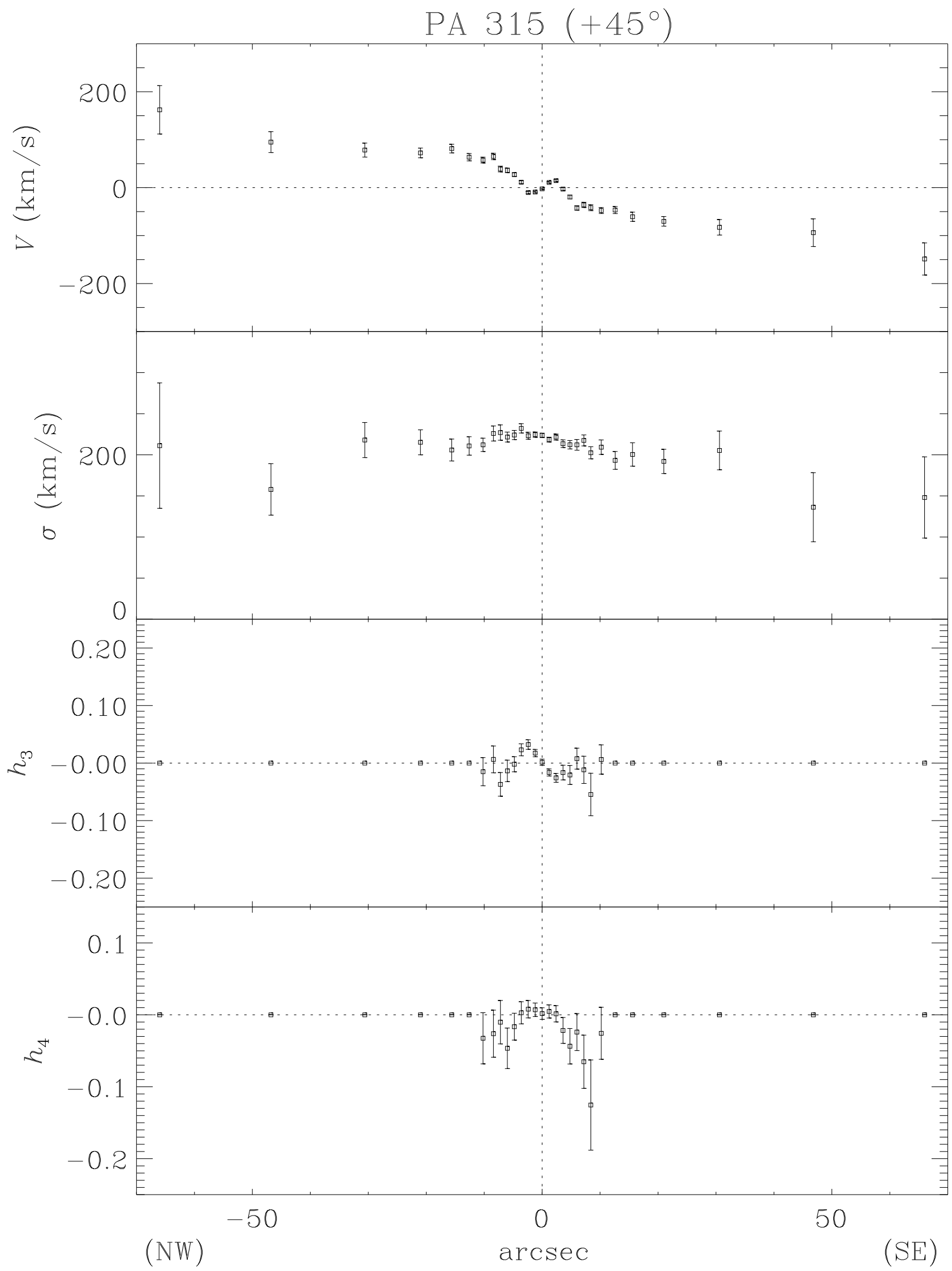

Fig. 8.— As in figure 5, but for the other diagonal slit position (PA 315).

The FCQ-derived velocity distributions are plotted as dotted lines in figure 9. In the end, no filtering strategy was able to eliminate the ringing around zero on either side of the peak, and we have had to add a constant vertical offset to all of the curves to line them up with the parametrized profiles. Having done so,

TABLE 2A. Data for PA 0

| $R('')$ | $V$ | ± | $\sigma$ | ± | $h_3$ | ± | $h_4$ | ± | $\langle v\rangle$ | ± | $\langle(v-\langle v\rangle)^2\rangle^{1/2}$ | ± |
|---|---|---|---|---|---|---|---|---|---|---|---|---|
| −66.0 | — | — | — | — | — | — | — | — | — | — | — | — |
| −46.8 | −8.0 | 36.2 | 174.6 | 53.6 | *−0.390* | *0.164* | *−0.175* | *0.405* | −8.0 | 36.2 | 174.6 | 53.6 |
| −30.6 | −16.5 | 16.2 | 235.9 | 24.7 | *0.048* | *0.059* | *−0.027* | *0.086* | −16.5 | 16.2 | 235.9 | 24.7 |
| −21.0 | 2.0 | 9.7 | 189.5 | 14.2 | *−0.004* | *0.045* | *−0.023* | *0.064* | 2.0 | 9.7 | 189.5 | 14.2 |
| −15.6 | −3.5 | 9.7 | 213.6 | 14.4 | *0.022* | *0.044* | *−0.069* | *0.071* | −3.5 | 9.7 | 213.6 | 14.4 |
| −12.6 | 8.0 | 7.9 | 200.6 | 11.7 | *−0.019* | *0.040* | *−0.055* | *0.064* | 8.0 | 7.9 | 200.6 | 11.7 |
| −10.2 | 0.5 | 7.8 | 231.1 | 12.8 | 0.039 | 0.026 | −0.012 | 0.041 | 14.4 | 14.8 | 226.6 | 25.7 |
| −8.4 | 10.7 | 6.7 | 222.3 | 11.7 | 0.036 | 0.023 | 0.008 | 0.034 | 24.3 | 12.1 | 226.1 | 25.4 |
| −7.2 | 16.8 | 5.6 | 203.8 | 9.6 | 0.026 | 0.022 | 0.009 | 0.033 | 25.8 | 10.1 | 208.0 | 22.2 |
| −6.0 | 4.9 | 4.8 | 195.3 | 5.9 | −0.008 | 0.025 | −0.065 | 0.041 | 3.5 | 6.1 | 178.6 | 7.8 |
| −4.8 | 2.7 | 3.8 | 222.0 | 6.4 | −0.008 | 0.014 | −0.005 | 0.020 | −0.2 | 6.9 | 219.3 | 14.7 |
| −3.6 | 5.0 | 3.1 | 226.3 | 5.4 | −0.010 | 0.011 | −0.005 | 0.016 | 1.3 | 5.8 | 223.3 | 12.4 |
| −2.4 | 1.8 | 2.3 | 220.4 | 4.5 | −0.013 | 0.008 | 0.017 | 0.012 | −3.3 | 4.3 | 229.1 | 9.7 |
| −1.2 | −3.7 | 1.7 | 224.3 | 3.1 | −0.012 | 0.006 | 0.004 | 0.009 | −8.2 | 3.1 | 226.1 | 7.1 |
| 0.0 | −3.7 | 1.4 | 229.5 | 2.7 | −0.006 | 0.005 | 0.007 | 0.007 | −6.2 | 2.6 | 233.5 | 6.0 |
| 1.2 | −6.0 | 1.6 | 225.0 | 3.0 | −0.004 | 0.006 | 0.009 | 0.008 | −7.5 | 3.0 | 229.8 | 6.7 |
| 2.4 | −3.7 | 2.2 | 217.8 | 4.1 | −0.005 | 0.008 | 0.013 | 0.012 | −5.5 | 4.1 | 224.8 | 9.2 |
| 3.6 | 8.6 | 3.0 | 230.0 | 8.1 | 0.007 | 0.010 | 0.048 | 0.016 | 11.2 | 5.8 | 255.0 | 15.3 |
| 4.8 | 1.7 | 3.5 | 204.7 | 5.5 | −0.007 | 0.014 | −0.009 | 0.021 | −0.8 | 6.2 | 200.3 | 12.4 |
| 6.0 | 1.4 | 4.5 | 217.0 | 6.1 | −0.021 | 0.018 | −0.032 | 0.025 | −4.4 | 6.8 | 204.9 | 9.9 |
| 7.2 | 3.1 | 5.2 | 217.5 | 7.0 | 0.013 | 0.020 | −0.032 | 0.029 | 6.8 | 7.7 | 205.0 | 11.4 |
| 8.4 | −1.1 | 5.6 | 205.5 | 6.9 | −0.024 | 0.026 | −0.056 | 0.041 | −5.9 | 7.4 | 189.6 | 9.7 |
| 10.2 | 1.5 | 5.6 | 190.8 | 6.6 | −0.014 | 0.033 | −0.092 | 0.056 | −0.4 | 6.7 | 172.5 | 7.8 |
| 12.6 | 18.3 | 7.6 | 206.8 | 11.2 | *−0.061* | *0.048* | *−0.159* | *0.081* | 18.3 | 7.6 | 206.8 | 11.2 |
| 15.6 | 5.6 | 8.7 | 178.7 | 12.6 | *−0.056* | *0.058* | *−0.059* | *0.084* | 5.6 | 8.7 | 178.7 | 12.6 |
| 21.0 | −21.4 | 8.9 | 175.7 | 13.0 | *0.054* | *0.062* | *−0.111* | *0.102* | −21.4 | 8.9 | 175.7 | 13.0 |
| 30.6 | 1.9 | 17.3 | 195.8 | 24.5 | *−0.026* | *0.108* | *−0.132* | *0.187* | 1.9 | 17.3 | 195.8 | 24.5 |
| 46.8 | 16.6 | 55.5 | 209.4 | 79.2 | *0.039* | *0.354* | *−0.505* | *0.583* | 16.6 | 55.5 | 209.4 | 79.2 |
| 66.0 | — | — | — | — | — | — | — | — | — | — | — | — |

Table 2B. Data for PA 225

| $R('')$ | $V$ | ± | $\sigma$ | ± | $h_3$ | ± | $h_4$ | ± | $\langle v \rangle$ | ± | $\langle (v - \langle v \rangle)^2 \rangle^{1/2}$ | ± |
|---|---|---|---|---|---|---|---|---|---|---|---|---|
| −66.0 | 155.9 | 65.0 | 145.0 | 95.3 | *−0.172* | *0.318* | *0.282* | *0.340* | 155.9 | 65.0 | 145.0 | 95.3 |
| −46.8 | 84.7 | 29.3 | 181.9 | 43.3 | *0.399* | *0.147* | *−0.168* | *0.426* | 84.7 | 29.3 | 181.9 | 43.3 |
| −30.6 | 71.2 | 15.4 | 188.5 | 22.9 | *0.136* | *0.068* | *0.122* | *0.057* | 71.2 | 15.4 | 188.5 | 22.9 |
| −21.0 | 71.4 | 10.3 | 196.0 | 15.1 | *0.006* | *0.042* | *0.037* | *0.059* | 71.4 | 10.3 | 196.0 | 15.1 |
| −15.6 | 74.1 | 8.4 | 184.6 | 12.2 | *0.169* | *0.071* | *0.165* | *0.034* | 74.1 | 8.4 | 184.6 | 12.2 |
| −12.6 | 52.4 | 7.8 | 217.0 | 11.6 | *−0.068* | *0.037* | *−0.050* | *0.050* | 52.4 | 7.8 | 217.0 | 11.6 |
| −10.2 | 66.3 | 6.4 | 204.7 | 8.8 | −0.019 | 0.026 | −0.022 | 0.037 | 60.9 | 10.2 | 195.8 | 16.3 |
| −8.4 | 56.6 | 6.3 | 225.8 | 8.2 | −0.037 | 0.024 | −0.044 | 0.035 | 47.4 | 9.2 | 211.3 | 12.3 |
| −7.2 | 52.6 | 5.2 | 225.8 | 6.7 | −0.021 | 0.020 | −0.045 | 0.029 | 47.4 | 7.3 | 210.2 | 9.7 |
| −6.0 | 33.3 | 4.3 | 218.2 | 7.9 | 0.024 | 0.016 | 0.011 | 0.023 | 42.1 | 8.0 | 223.9 | 17.9 |
| −4.8 | 17.8 | 3.9 | 232.8 | 6.6 | −0.009 | 0.013 | −0.010 | 0.019 | 14.3 | 7.1 | 227.6 | 14.2 |
| −3.6 | 11.5 | 3.1 | 228.1 | 6.0 | 0.040 | 0.010 | 0.015 | 0.016 | 27.3 | 5.7 | 236.0 | 13.0 |
| −2.4 | 0.7 | 2.4 | 224.9 | 4.8 | 0.011 | 0.009 | 0.016 | 0.013 | 4.8 | 4.5 | 233.3 | 10.6 |
| −1.2 | −2.3 | 1.8 | 226.6 | 3.4 | −0.010 | 0.006 | 0.007 | 0.010 | −6.4 | 3.4 | 230.2 | 7.8 |
| 0.0 | 8.5 | 1.7 | 222.0 | 2.7 | −0.027 | 0.006 | −0.004 | 0.009 | −1.5 | 3.0 | 220.0 | 5.7 |
| 1.2 | 18.2 | 1.9 | 221.8 | 3.2 | −0.037 | 0.007 | 0.005 | 0.009 | 4.2 | 3.3 | 224.4 | 6.8 |
| 2.4 | 16.0 | 2.4 | 224.5 | 4.2 | −0.030 | 0.008 | 0.003 | 0.012 | 4.4 | 4.3 | 226.1 | 9.1 |
| 3.6 | −3.2 | 3.3 | 228.8 | 5.4 | −0.021 | 0.011 | −0.006 | 0.016 | −11.4 | 5.9 | 225.7 | 11.4 |
| 4.8 | −23.6 | 3.9 | 227.3 | 5.0 | −0.000 | 0.014 | −0.041 | 0.021 | −23.7 | 5.4 | 211.8 | 7.4 |
| 6.0 | −42.1 | 4.5 | 213.4 | 5.8 | −0.025 | 0.019 | −0.047 | 0.029 | −47.7 | 6.2 | 198.3 | 8.4 |
| 7.2 | −56.2 | 5.6 | 232.1 | 8.1 | 0.011 | 0.019 | −0.022 | 0.026 | −52.5 | 9.1 | 221.6 | 14.4 |
| 8.4 | −59.0 | 6.4 | 218.4 | 8.3 | 0.018 | 0.025 | −0.039 | 0.037 | −54.4 | 9.2 | 204.5 | 12.8 |
| 10.2 | −62.1 | 6.8 | 186.1 | 8.7 | 0.102 | 0.053 | −0.110 | 0.073 | −49.7 | 8.1 | 171.2 | 10.0 |
| 12.6 | −63.0 | 7.8 | 189.7 | 11.5 | *0.047* | *0.035* | *0.003* | *0.051* | −63.0 | 7.8 | 189.7 | 11.5 |
| 15.6 | −72.9 | 7.9 | 178.5 | 11.7 | *0.031* | *0.168* | *0.207* | *0.025* | −72.9 | 7.9 | 178.5 | 11.7 |
| 21.0 | −85.0 | 9.0 | 172.3 | 13.1 | *0.094* | *0.045* | *0.154* | *0.060* | −85.0 | 9.0 | 172.3 | 13.1 |
| 30.6 | −83.3 | 13.8 | 156.1 | 20.2 | *0.287* | *0.207* | *0.182* | *0.110* | −83.3 | 13.8 | 156.1 | 20.2 |
| 46.8 | −31.9 | 24.1 | 187.1 | 34.6 | *−0.360* | *0.264* | *−0.115* | *0.294* | −31.9 | 24.1 | 187.1 | 34.6 |
| 66.0 | 234.8 | 108.9 | 35.8 | 228.9 | *−0.467* | *0.397* | *0.570* | *0.439* | 234.8 | 108.9 | 35.8 | 202.6 |

Table 2C. Data for PA 270

| $R('')$ | $V$ | ± | $\sigma$ | ± | $h_3$ | ± | $h_4$ | ± | $\langle v \rangle$ | ± | $\langle (v - \langle v \rangle)^2 \rangle^{1/2}$ | ± |
|---|---|---|---|---|---|---|---|---|---|---|---|---|
| −66.0 | 92.9 | 25.4 | 85.5 | 38.3 | *−0.249* | *1.605* | *−1.203* | *3.735* | 92.9 | 25.4 | 85.5 | 38.3 |
| −46.8 | 140.6 | 17.9 | 115.9 | 26.5 | *0.160* | *0.415* | *0.343* | *0.105* | 140.6 | 17.9 | 115.9 | 26.5 |
| −30.6 | 132.7 | 11.4 | 184.7 | 16.6 | *0.142* | *0.117* | *−0.313* | *0.193* | 132.7 | 11.4 | 184.7 | 16.6 |
| −21.0 | 114.0 | 9.2 | 194.8 | 13.3 | *−0.281* | *0.088* | *−0.293* | *0.122* | 114.0 | 9.2 | 194.8 | 13.3 |
| −15.6 | 107.9 | 8.6 | 180.5 | 12.4 | *−0.015* | *0.040* | *0.029* | *0.057* | 107.9 | 8.6 | 180.5 | 12.4 |
| −12.6 | 90.2 | 8.5 | 210.2 | 12.4 | *−0.065* | *0.043* | *−0.065* | *0.061* | 90.2 | 8.5 | 210.2 | 12.4 |
| −10.2 | 86.8 | 8.1 | 213.9 | 11.1 | −0.073 | 0.034 | −0.028 | 0.049 | 66.8 | 13.7 | 208.2 | 19.6 |
| −8.4 | 74.6 | 9.5 | 234.0 | 10.3 | −0.152 | 0.042 | −0.127 | 0.054 | 53.0 | 11.1 | 218.9 | 11.3 |
| −7.2 | 75.2 | 7.9 | 232.2 | 8.7 | −0.099 | 0.039 | −0.173 | 0.063 | 65.0 | 8.6 | 209.4 | 8.5 |
| −6.0 | 42.8 | 6.3 | 226.8 | 7.0 | −0.059 | 0.029 | −0.116 | 0.046 | 34.7 | 7.2 | 205.0 | 7.7 |
| −4.8 | 36.7 | 5.4 | 247.1 | 6.5 | −0.016 | 0.018 | −0.062 | 0.028 | 33.0 | 7.0 | 226.7 | 8.3 |
| −3.6 | 2.5 | 4.4 | 240.5 | 5.6 | 0.054 | 0.015 | −0.052 | 0.022 | 15.7 | 6.4 | 225.0 | 8.0 |
| −2.4 | −26.6 | 3.0 | 222.8 | 5.7 | 0.029 | 0.011 | 0.013 | 0.016 | −15.5 | 5.6 | 229.4 | 12.6 |
| −1.2 | −22.9 | 2.2 | 220.8 | 4.2 | 0.018 | 0.008 | 0.014 | 0.012 | −16.2 | 4.1 | 228.3 | 9.1 |
| 0.0 | 0.7 | 1.9 | 221.4 | 3.1 | −0.006 | 0.007 | −0.013 | 0.010 | −1.6 | 3.4 | 215.0 | 6.5 |
| 1.2 | 19.0 | 2.2 | 223.3 | 3.9 | −0.027 | 0.008 | 0.004 | 0.011 | 8.6 | 4.1 | 225.2 | 8.7 |
| 2.4 | 15.8 | 3.6 | 237.0 | 7.2 | −0.057 | 0.011 | 0.026 | 0.016 | −7.3 | 6.5 | 250.5 | 14.5 |
| 3.6 | 4.0 | 4.8 | 254.3 | 8.0 | −0.022 | 0.014 | −0.012 | 0.020 | −4.8 | 8.8 | 247.7 | 15.9 |
| 4.8 | −30.2 | 5.2 | 232.9 | 6.1 | −0.018 | 0.023 | −0.110 | 0.038 | −32.8 | 6.1 | 209.3 | 6.7 |
| 6.0 | −47.9 | 6.5 | 246.6 | 8.9 | −0.011 | 0.021 | −0.034 | 0.029 | −51.2 | 9.7 | 231.7 | 13.6 |
| 7.2 | −67.8 | 8.7 | 234.0 | 10.2 | 0.149 | 0.041 | −0.110 | 0.050 | −44.3 | 10.8 | 220.7 | 11.8 |
| 8.4 | −61.0 | 8.5 | 244.1 | 10.0 | 0.065 | 0.031 | −0.081 | 0.045 | −48.4 | 10.9 | 224.4 | 12.1 |
| 10.2 | −82.2 | 7.4 | 203.0 | 9.1 | 0.118 | 0.049 | −0.130 | 0.068 | −68.4 | 8.5 | 186.5 | 9.7 |
| 12.6 | −99.4 | 8.4 | 210.8 | 12.5 | *0.115* | *0.051* | *−0.061* | *0.064* | −99.4 | 8.4 | 210.8 | 12.5 |
| 15.6 | −100.0 | 8.7 | 208.1 | 12.8 | *−0.013* | *0.043* | *−0.064* | *0.070* | −100.0 | 8.7 | 208.1 | 12.8 |
| 21.0 | −113.4 | 10.2 | 200.7 | 14.9 | *0.118* | *0.069* | *−0.152* | *0.103* | −113.4 | 10.2 | 200.7 | 14.9 |
| 30.6 | −134.2 | 19.6 | 273.9 | 30.0 | *0.037* | *0.068* | *0.252* | *0.081* | −134.2 | 19.6 | 273.9 | 30.0 |
| 46.8 | −118.3 | 27.1 | 199.9 | 40.7 | *−0.296* | *0.983* | *0.193* | *0.434* | −118.3 | 27.1 | 199.9 | 40.7 |
| 66.0 | −36.5 | 52.4 | 158.4 | 77.6 | *0.298* | *0.968* | *0.838* | *3.342* | −36.5 | 52.4 | 158.4 | 77.6 |

Table 2D. Data for PA 315

| $R('')$ | $V$ | $\pm$ | $\sigma$ | $\pm$ | $h_3$ | $\pm$ | $h_4$ | $\pm$ | $\langle v\rangle$ | $\pm$ | $\langle(v-\langle v\rangle)^2\rangle^{1/2}$ | $\pm$ |
|---|---|---|---|---|---|---|---|---|---|---|---|---|
| −66.0 | 162.3 | 50.5 | 211.1 | 76.3 | *0.747* | *1.614* | *0.143* | *2.573* | 162.3 | 50.5 | 211.1 | 76.3 |
| −46.8 | 94.9 | 21.8 | 157.9 | 31.3 | *0.045* | *0.224* | *−0.192* | *0.375* | 94.9 | 21.8 | 157.9 | 31.3 |
| −30.6 | 78.3 | 14.6 | 218.0 | 21.4 | *−0.027* | *0.083* | *−0.179* | *0.146* | 78.3 | 14.6 | 218.0 | 21.4 |
| −21.0 | 72.4 | 10.2 | 215.2 | 15.2 | *−0.049* | *0.053* | *−0.079* | *0.078* | 72.4 | 10.2 | 215.2 | 15.2 |
| −15.6 | 81.5 | 9.0 | 205.9 | 13.3 | *−0.010* | *0.037* | *−0.017* | *0.052* | 81.5 | 9.0 | 205.9 | 13.3 |
| −12.6 | 63.5 | 7.6 | 210.8 | 11.2 | *−0.053* | *0.029* | *0.046* | *0.039* | 63.5 | 7.6 | 210.8 | 11.2 |
| −10.2 | 57.4 | 6.1 | 212.2 | 8.1 | −0.015 | 0.025 | −0.033 | 0.036 | 53.5 | 9.0 | 199.8 | 13.2 |
| −8.4 | 65.1 | 6.6 | 225.9 | 9.1 | 0.006 | 0.023 | −0.026 | 0.033 | 67.0 | 10.2 | 214.4 | 15.5 |
| −7.2 | 38.9 | 6.0 | 227.1 | 9.3 | −0.037 | 0.021 | −0.010 | 0.030 | 25.6 | 10.8 | 223.0 | 18.2 |
| −6.0 | 36.0 | 4.7 | 221.6 | 6.1 | −0.014 | 0.019 | −0.047 | 0.028 | 32.8 | 6.5 | 205.7 | 8.8 |
| −4.8 | 27.3 | 3.7 | 224.1 | 5.6 | −0.002 | 0.013 | −0.016 | 0.019 | 26.6 | 6.2 | 215.9 | 11.0 |
| −3.6 | 11.4 | 3.1 | 232.2 | 5.6 | 0.023 | 0.010 | 0.003 | 0.015 | 20.6 | 5.8 | 233.6 | 12.6 |
| −2.4 | −10.2 | 2.4 | 223.1 | 4.3 | 0.032 | 0.008 | 0.008 | 0.012 | 2.2 | 4.3 | 227.0 | 9.4 |
| −1.2 | −9.2 | 1.8 | 224.5 | 3.3 | 0.017 | 0.006 | 0.007 | 0.009 | −2.5 | 3.3 | 228.3 | 7.5 |
| 0.0 | −2.6 | 1.5 | 223.7 | 2.7 | 0.001 | 0.005 | 0.002 | 0.008 | −2.0 | 2.8 | 224.6 | 6.4 |
| 1.2 | 10.8 | 1.7 | 218.5 | 3.0 | −0.016 | 0.006 | 0.005 | 0.009 | 4.6 | 3.1 | 220.9 | 6.9 |
| 2.4 | 14.6 | 2.2 | 221.6 | 3.7 | −0.026 | 0.008 | 0.001 | 0.011 | 4.8 | 3.9 | 222.3 | 8.2 |
| 3.6 | −3.0 | 3.3 | 213.5 | 4.7 | −0.016 | 0.013 | −0.022 | 0.018 | −8.0 | 5.3 | 204.3 | 8.6 |
| 4.8 | −19.5 | 4.0 | 212.3 | 5.1 | −0.021 | 0.017 | −0.044 | 0.025 | −24.3 | 5.6 | 197.8 | 7.6 |
| 6.0 | −42.5 | 4.7 | 212.2 | 6.5 | 0.008 | 0.018 | −0.024 | 0.026 | −40.2 | 7.3 | 202.0 | 11.5 |
| 7.2 | −36.0 | 5.5 | 217.5 | 6.7 | −0.012 | 0.024 | −0.065 | 0.037 | −38.2 | 7.0 | 199.1 | 8.6 |
| 8.4 | −41.8 | 6.3 | 202.5 | 7.3 | −0.055 | 0.037 | −0.125 | 0.063 | −48.1 | 7.3 | 182.3 | 8.1 |
| 10.2 | −47.7 | 6.4 | 209.3 | 8.8 | 0.006 | 0.025 | −0.026 | 0.036 | −45.9 | 9.7 | 198.7 | 15.2 |
| 12.6 | −46.8 | 7.4 | 193.2 | 10.8 | *−0.015* | *0.035* | *0.144* | *0.027* | −46.8 | 7.4 | 193.2 | 10.8 |
| 15.6 | −60.7 | 9.6 | 200.4 | 14.2 | *0.034* | *0.038* | *0.101* | *0.052* | −60.7 | 9.6 | 200.4 | 14.2 |
| 21.0 | −70.3 | 10.0 | 192.0 | 14.7 | *0.112* | *0.045* | *0.036* | *0.062* | −70.3 | 10.0 | 192.0 | 14.7 |
| 30.6 | −82.7 | 16.2 | 205.3 | 23.6 | *0.233* | *0.169* | *−0.470* | *0.253* | −82.7 | 16.2 | 205.3 | 23.6 |
| 46.8 | −93.9 | 28.8 | 136.2 | 42.0 | *−0.547* | *0.420* | *−0.352* | *0.753* | −93.9 | 28.8 | 136.2 | 42.0 |
| 66.0 | −148.7 | 33.6 | 148.1 | 49.4 | *0.307* | *0.762* | *0.201* | *0.433* | −148.7 | 33.6 | 148.1 | 49.4 |

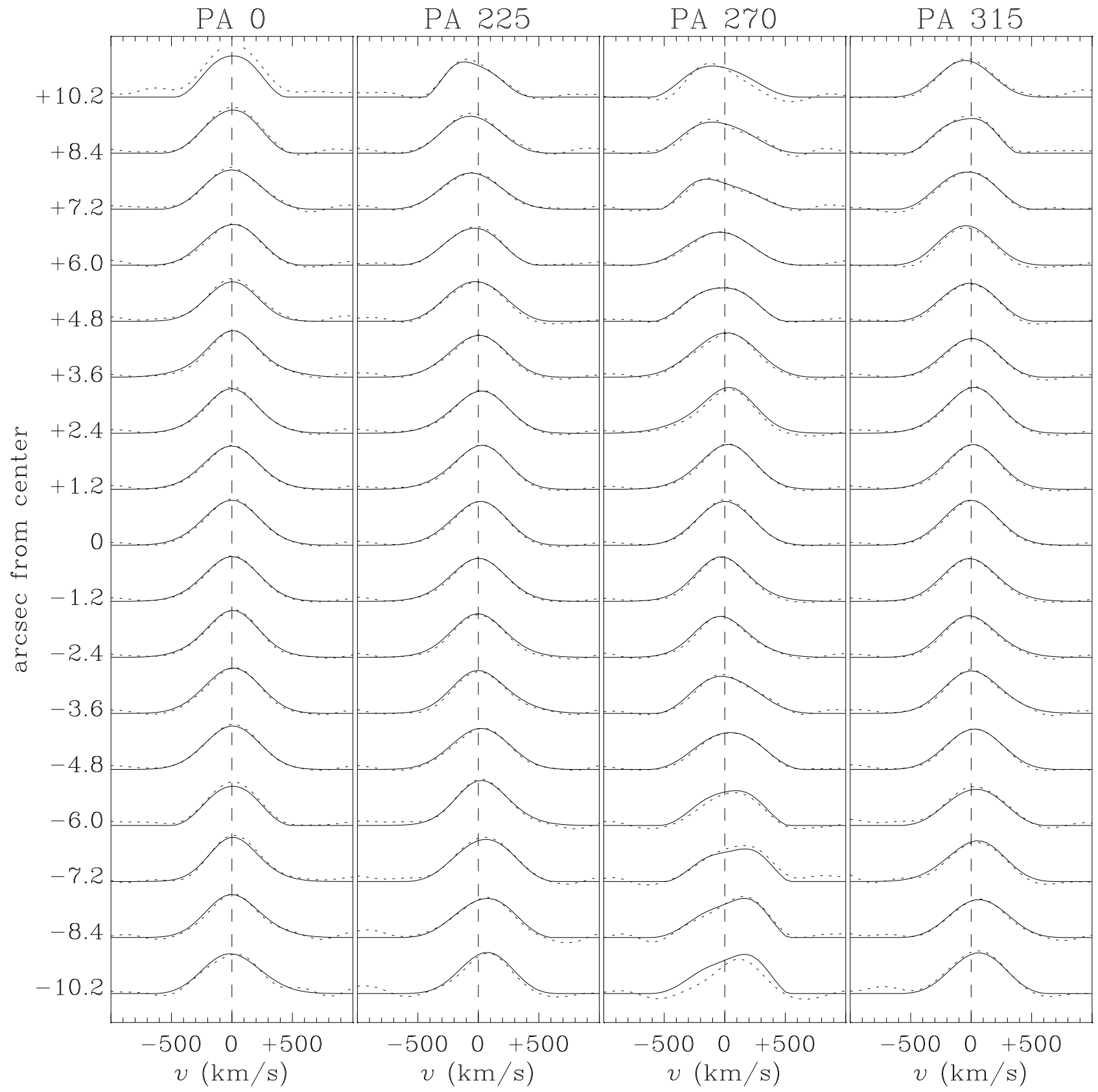

Fig. 9.— *Solid lines,* line-of-sight velocity distributions for the inner part of NGC 1700, reconstructed from the fitted values of $V$, $\sigma$, $h_3$, and $h_4$ obtained from the cross-correlation method. Note the near-Gaussian profile, even in the counterrotating core (clearest on PA 270). *Dotted lines,* Non-parametric LOSVDs derived using the Fourier correlation quotient method. The agreement is excellent.

however, the agreement is very good, showing that our results do not depend on choosing the parametric representation, equation (3). In particular, there is no sign of a separate rapidly rotating component in the core, as might be expected from a central disk.

### 3.3. Mean Velocity Field

The quantity $V$ plotted in figures 5 – 8 is not quite the true mean velocity, because of the $h_3$ and $h_4$ terms in equation (3). Similarly, $\sigma$ is not quite the true velocity dispersion. vdMF give expressions for the mean velocity $\langle v \rangle$ and dispersion $(\langle v^2 \rangle - \langle v \rangle^2)^{1/2}$ in terms of $V$, $\sigma$, $h_3$, and $h_4$, but their expressions pertain only to an unmodified Gauss-Hermite expansion and do not apply here because we are enforcing non-negativity of $L(v)$. Instead, we have

$$\langle v \rangle = V + \sigma M(h_3, h_4), \qquad \left(\langle v^2 \rangle - \langle v \rangle^2\right)^{1/2} = \sigma D(h_3, h_4), \tag{5}$$

where $M(h_3,h_4)$ and $D(h_3,h_4)$ are respectively the first moment and the square root of the second moment of $L(v)$ for $V=0$ and $\sigma=1$, and it is understood that the unit of velocity is $1\,\mathrm{km\,s^{-1}}$. We compute $M$, $D$, and their partial derivatives by numerical integration. The formal errors in the corrected quantities follow from, *e.g.*,

$$\sigma^2_{\langle v\rangle} = [d][C][d]^T, \tag{6}$$

where $[d]$ is the row matrix

$$[d] = \left[\begin{array}{ccccc} \frac{\partial\langle v\rangle}{\partial V} & \frac{\partial\langle v\rangle}{\partial \sigma} & \frac{\partial\langle v\rangle}{\partial \gamma} & \frac{\partial\langle v\rangle}{\partial h_3} & \frac{\partial\langle v\rangle}{\partial h_4} \end{array}\right] = \left[\begin{array}{ccccc} 1 & M & 0 & \sigma\frac{\partial M}{\partial h_3} & \sigma\frac{\partial M}{\partial h_4} \end{array}\right], \tag{7}$$

and $[C]$ is the covariance matrix of the fit (with the same ordering of parameters).

The corrected mean velocity profiles are shown in figure 10 and listed in column 10 of Table 2. Comparison of the third panel with figure 7 shows that the actual *mean* counterrotation in the core is very small, only $\sim 15\,\mathrm{km\,s^{-1}}$; most of the effect is due to the modest skewness gradient. This again argues against a nuclear disk as the source of the counterrotation. Elsewhere, corrections to the mean velocity field are not qualitatively significant. Out to $\sim 40''$ the profiles are strikingly antisymmetric, characteristic of a relaxed and well-mixed system. Beyond this point, note the turn-down of the rotation curve on PA 270 and the turn-up on PA 315, indicating a twist of the kinematic major axis (the position angle of maximum velocity at fixed $R$) away from the photometric major axis. This is a sign of either a radial gradient in the intrinsic axis ratios or a spatial twist of the isodensity surfaces.

Of particular interest is the velocity reversal at large positive radii (northeast) on PA 225. This sort of feature might be expected from the isophote twist and boxiness seen at still larger radii (Schweizer & Seitzer 1992), since velocity-reversed streamers are typical artifacts of tidal interactions (Borne & Hoessel 1988, Borne *et al.* 1994); but its appearance on this PA is surprising since most of the interesting photometric structures are seen around PA 135/315. Yet despite the systematic difficulties with this outermost bin (*cf.* Sec. 2.3.1), we believe the velocity reversal is real, for the following reasons: first, the amplitude is very large, and significant at the $2\sigma$ level even with our liberal estimates of the systematic errors in this bin; second, the next-innermost point shows the beginning of a turnover in the rotation curve; finally, the corresponding points on the other PAs are not anomalous. We are currently reducing spectra obtained at larger radii to further explore the velocity reversal, and the results will be discussed in a subsequent paper.

At a given projected radius $R$, we can create a Fourier reconstruction of the angular dependence of the velocity field. With eight angular samples at each radius, without artificially imposing symmetry across the center we have enough information for a representation of the form

$$\begin{aligned} v(R,\theta) = C_0(R) \;+\; & C_1(R)\cos\theta + S_1(R)\sin\theta + C_2(R)\cos 2\theta + S_2(R)\sin 2\theta \\ +\; & C_3(R)\cos 3\theta + S_3(R)\sin 3\theta + C_4(R)\cos 4\theta, \end{aligned} \tag{8}$$

where $\theta$ is the angle measured counterclockwise from the major axis. We cannot recover the complete fourth harmonic since $\sin 4\theta = 0$ on the sampled PAs. The velocity at arbitrary $(R,\theta)$ is then

$$v(R,\theta) = \left[\begin{array}{c} v(R,0^\circ) \\ v(R,45^\circ) \\ v(R,90^\circ) \\ v(R,135^\circ) \\ v(R,180^\circ) \\ v(R,225^\circ) \\ v(R,270^\circ) \\ v(R,315^\circ) \end{array}\right]^T \left[\begin{array}{cccccccc} 1 & 2 & 0 & 2 & 0 & 2 & 0 & 1 \\ 1 & \sqrt{2} & \sqrt{2} & 0 & 2 & -\sqrt{2} & \sqrt{2} & -1 \\ 1 & 0 & 2 & -2 & 0 & 0 & -2 & 1 \\ 1 & -\sqrt{2} & \sqrt{2} & 0 & -2 & \sqrt{2} & \sqrt{2} & -1 \\ 1 & -2 & 0 & 2 & 0 & -2 & 0 & 1 \\ 1 & -\sqrt{2} & -\sqrt{2} & 0 & 2 & \sqrt{2} & -\sqrt{2} & -1 \\ 1 & 0 & -2 & -2 & 0 & 0 & 2 & 1 \\ 1 & \sqrt{2} & -\sqrt{2} & 0 & -2 & -\sqrt{2} & -\sqrt{2} & -1 \end{array}\right] \left[\begin{array}{c} 1 \\ \cos\theta \\ \sin\theta \\ \cos 2\theta \\ \sin 2\theta \\ \cos 3\theta \\ \sin 3\theta \\ \cos 4\theta \end{array}\right], \tag{9}$$

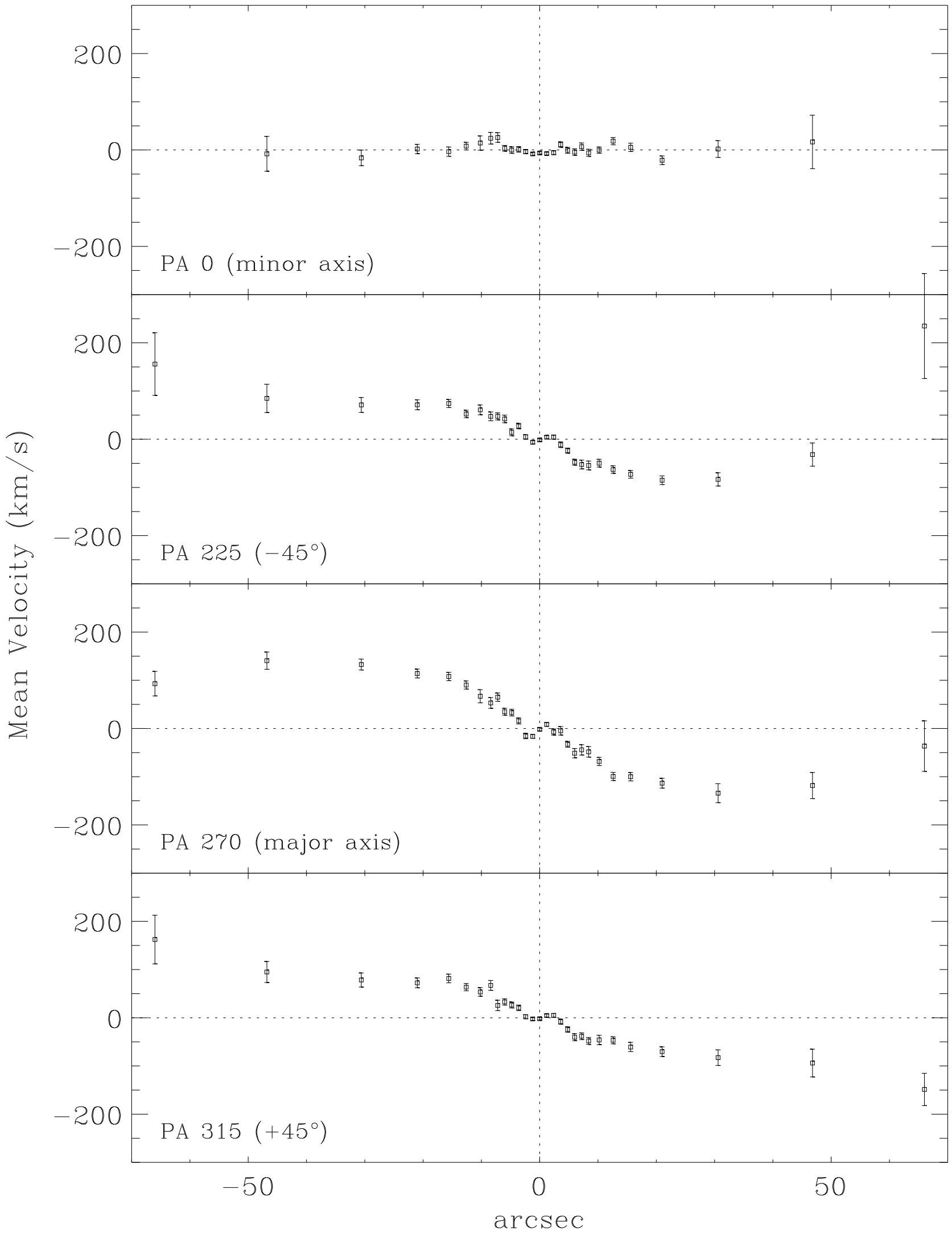

Fig. 10.— Mean velocity profiles, corrected for the non-Gaussian terms as described in sec. 3.3. Note the reduced amplitude of counterrotation in the core.

where the velocities in the leading column matrix are obtained by linearly interpolating in $R$ between the values in column 10 of Table 2.

The Fourier-reconstructed velocity field for $R \leq 56''$ is shown in figure 11, overplotted with the fitted isophotes derived by Franx *et al.* (1989*a*). The important points to notice are (1) the regularity of the velocity field for $R \lesssim 35''$, with symmetry across the major axis and antisymmetry across the minor axis characteristic of near-oblateness; (2) the counterclockwise twist of the kinematic major axis for $R \gtrsim 35''$, accompanied by a mild isophotal twist in the same direction but *not* by a twist of the zero-velocity contour;

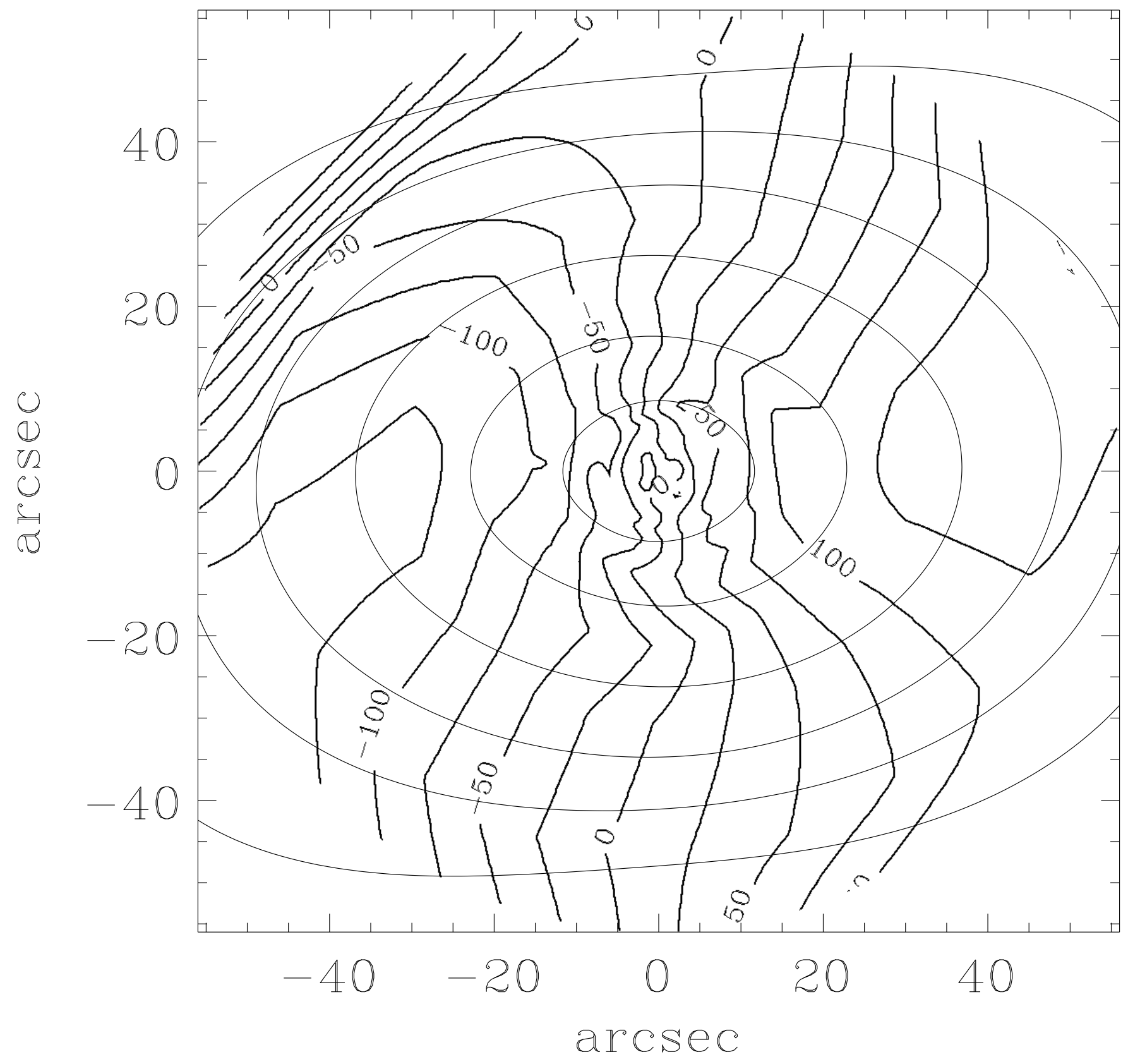

Fig. 11.— Fourier reconstruction of the 2-dimensional mean velocity field of NGC 1700, overplotted with fitted isophotes from Franx *et al.* (1989). Velocity contours are plotted every $25\,\mathrm{km\,s^{-1}}$. Note the counterrotating core, the counterclockwise twist of the kinematic major axis outside $\sim 35''$, and the velocity reversal at large radii to the northeast.

(3) the velocity reversal to the northeast. Each of these points is a clue to the galaxy's present configuration and past history, and will be further discussed in Sec. 4.

### 3.4. Velocity Dispersion Field

The corrected velocity dispersion profiles are shown in figure 12 and listed in column 12 of Table 2. The data have been folded about the center and averaged, and are plotted in logarithmic coordinates. The radial decrease in dispersion on each PA is consistent with a power law; the apparent steepening of the profile around $30''$ on the major axis is of marginal significance. A least-squares fit of a straight line through the data points gives the logarithmic gradient, $d\log\sigma/d\log R$. There are statistically significant, but small, differences in the slopes: we find $d\log\sigma/d\log R = -0.062 \pm 0.009$ on PA 0, $-0.075 \pm 0.009$ on PA 225, $-0.052 \pm 0.010$ on PA 270, and $-0.048 \pm 0.008$ on PA 315. These values are completely consistent with the average value of $-0.06$, with a cosmic scatter of about 0.03, found for normal ellipticals and brightest cluster galaxies by Fisher *et al.* (1995). Whether the differences in the dispersion gradients on the sampled PAs indicate something important about the galaxy's history is an open question, though it may be suggestive that the shallowest gradient is found in the direction of the excess light at large $R$ and the steepest in the orthogonal direction.

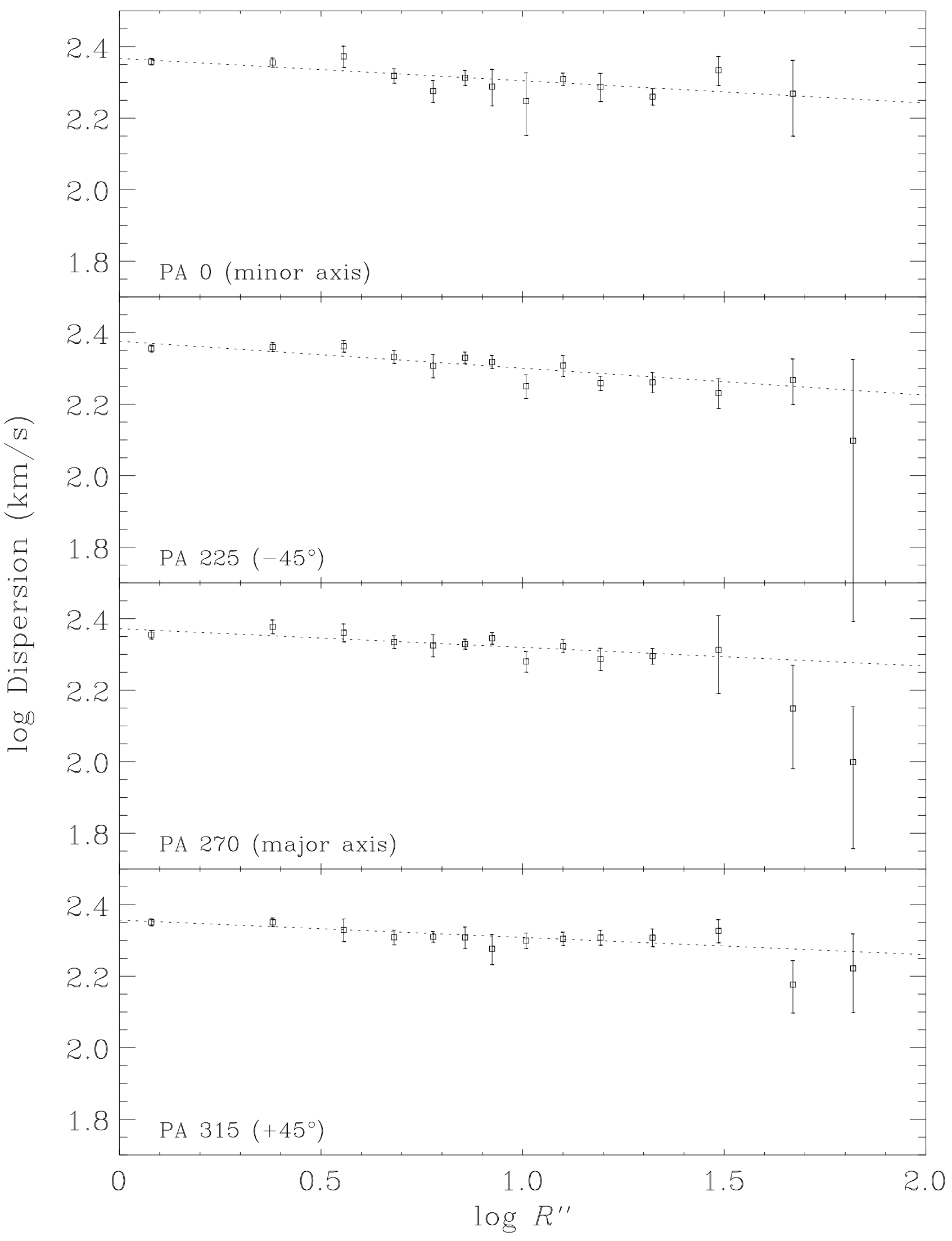

Fig. 12.— Velocity dispersion profiles, corrected for the non-Gaussian terms as described in sec. 3.3. Dotted lines show the best-fit power laws, with slopes given in sec. 3.4.

Dynamical support of the system is provided by a combination of rotation and random motions. The increasing importance of rotational support at larger radii is evident by comparing figure 12 with figure 13, where we plot the major-axis root-mean-square velocity profile. The logarithmic slope is $-0.012 \pm 0.009$, much flatter than the dispersion profile, and consistent with the expected behavior if NGC 1700 is embedded in a standard isothermal dark halo.

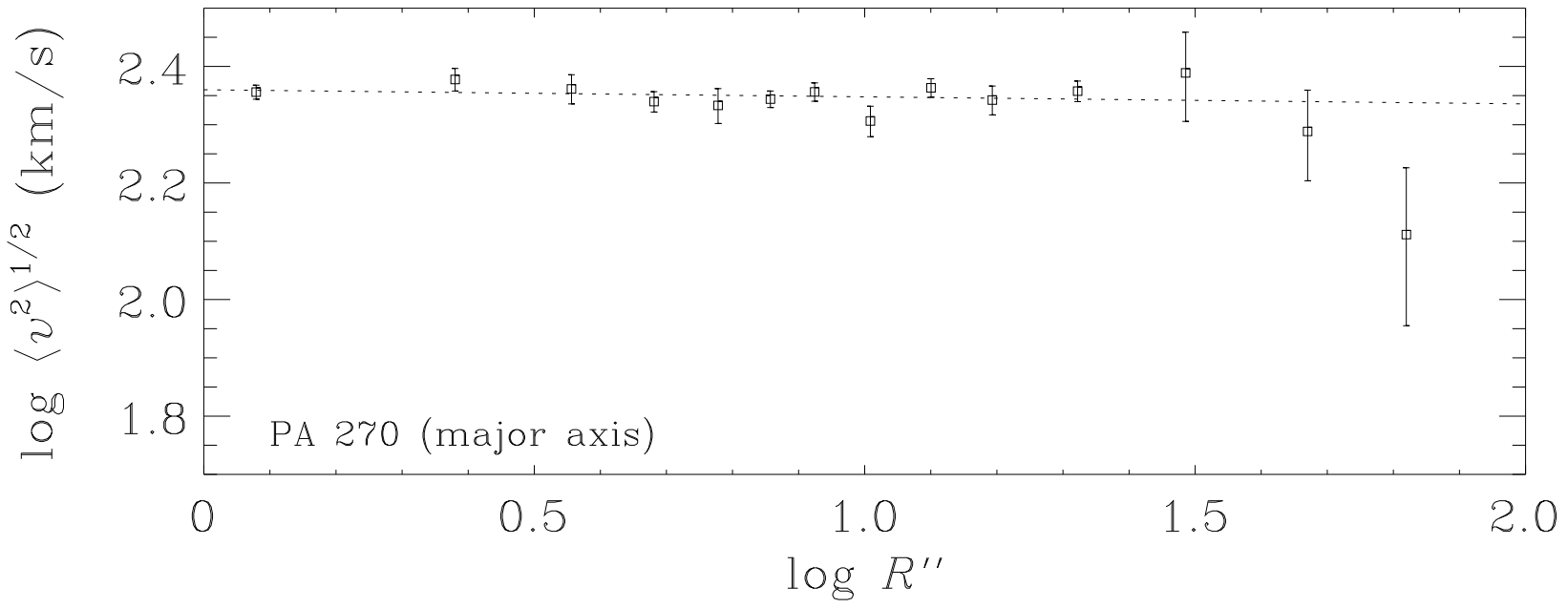

Fig. 13.— Major axis profile of the RMS projected velocity (*i.e.,* both rotational and random motion). Dotted line gives the best-fit power law, with a logarithmic slope of $-0.012 \pm 0.009$. This is significantly flatter than the dispersion profile, and consistent with a standard dark halo.

## 4. Discussion

### 4.1. Intrinsic Shape and Mass Distribution

We defer a full discussion of these issues to a later paper; but qualitative statements about the shape of NGC 1700 and the radial mass profile can be made from inspection of the kinematic fields.

The regularity of the velocity field in the main body ($5'' \lesssim R \lesssim 50''$) of the galaxy is highlighted in figure 14, in which we plot radial profiles of the three VF symmetry parameters defined by Statler & Fry (1994). The parameter $V_1$ is $2/\pi$ times the kinematic misalignment angle $\Psi$ (Franx *et al.* 1991), and approximately measures the orientation of the zero-velocity contour relative to the photometric minor axis; $V_2$ and $V_3$ measure the deviation of the VF's angular dependence from the simple form $v \propto \cos\theta$.[7] For axisymmetric oblate galaxies rotating about their symmetry axes, $V_1 \equiv 0$ and $|V_2|, |V_3| < 0.1$. For nearly oblate systems, $|V_1|$, $|V_2|$, and $|V_3|$ may take on larger values if the true rotation axis is intrinsically misaligned with the true minor axis; this occurs if there is net circulation in both short-axis and long-axis tube orbits. The data are consistent with completely flat profiles, $V_1 = V_2 = V_3 = 0$, showing that the velocity field is nearly self-similar over a factor $\sim 10$ in radius.

Statler & Fry (1994) demonstrate that the $V_i$ parameters can be used as approximate indicators of intrinsic shape, and present "channel maps" of the parameter space with which the most likely shape can be estimated from $V_1$, $V_2$, $V_3$, and the apparent ellipticity $\epsilon$. It can be seen from their figures 5–14 that, for $\epsilon \approx 0.3$, values of all three $V_i$'s near zero strongly favor small triaxiality $T$, regardless of the intrinsic rotational misalignment. If the misalignment is small, then $V_2$ gives the strongest constraint on $T$; if it is large, then $V_1$ and $V_3$ are more relevant. The flattening is harder to constrain, but if there is appreciable intrinsic misalignment, then for a given intrinsic short-to-long axis ratio $c/a$, small values of $|V_2|$ are more frequently seen in orientations where the galaxy looks round. Thus there is weak indication that NGC 1700 is substantially flatter than it appears.

The RMS velocity profile in figure 13 does suggest that NGC 1700 resides in a standard isothermal dark halo; but establishing the presence of dark matter in ellipticals is much harder than in spirals, since stellar dynamical information never extends as far as H I rotation curves and the orbits of the stars are not known *a priori*. Here the flat profile clearly extends at least to $3\,r_e$; but in a pure de Vaucouleurs law this radius encloses only 79% of the total light, so we are still not in the asymptotic large-$r$ regime. For the mass distribution to be isothermal-like requires only a factor $\sim$ 3 increase in $M/L$ from $1\,r_e$ to $3\,r_e$, and stellar population gradients could easily account for part of this. Alternatively, constant $M/L$ models can also show flat dispersion profiles, if there is a trend toward tangential anisotropy in the outer parts. Carter *et al.* (1985) and Bertin *et al.* (1994) both find that centrally isotropic models can mimic the dispersion profiles of isothermal spheres if the ratio of tangential to radial dispersion increases outward by a factor 2.5 – 3. Because most formation scenarios tend to favor increasing *radial* anisotropy at large $r$, such configurations are not widely discussed in the literature, but may be possible in mergers involving initially grazing encounters (Barnes 1992). Thus the evidence for dark matter in NGC 1700 should best be described as indicative, but not definitive.

[7]The $V_i$ are defined in terms of measured velocities on the principal axes and at the $\pm 45°$ diagonal PAs. NGC 1700's isophote twist thus makes our measured $V_i$ values somewhat fictitious in the outermost radial bin.

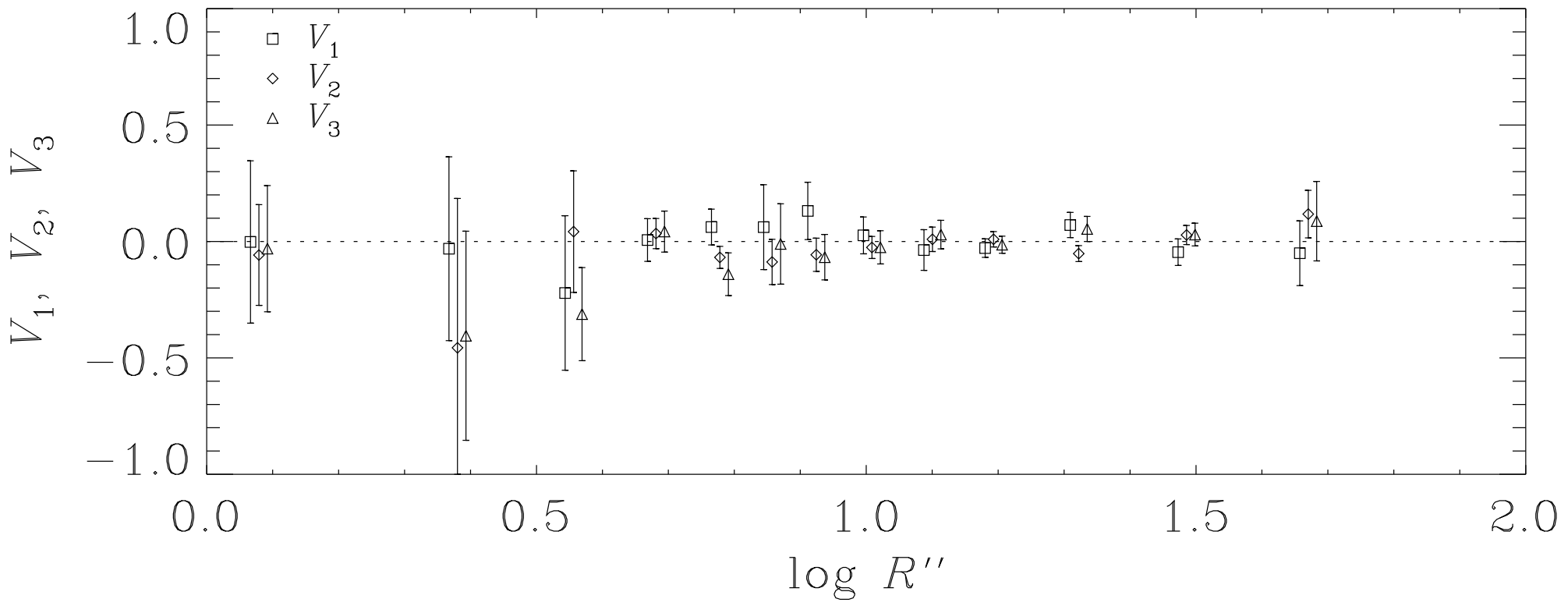

Fig. 14.— Radial profiles of the $V_1$, $V_2$, and $V_3$ parameters describing misalignment and asymmetry of the velocity field. Allowed values for each parameter span the range $[-1, 1]$, though $V_2$ and $V_3$ generally lie between $\pm 0.5$ for realistic systems. Axisymmetric oblate galaxies should have $V_1 \equiv 0$ and $V_2 \approx V_3 \approx 0$.

### 4.2. Counterrotating Core

Several possible origins have been proposed for counterrotation in E galaxy cores (Bender 1990*b* and references therein), the frontrunners being: (1) the remnant of an accreted stellar companion, initially orbiting contrary to the mean rotation of the galaxy (Kormendy 1984, Balcells & Quinn 1990); (2) a central stellar disk formed from accreted gas also with an initial counterrotating orbit (Franx & Illingworth 1988); (3) a central disk formed during a major merger of two spirals (Schweizer 1990); and (4) geometrical projection of the stellar orbits in the core of an otherwise ordinary triaxial galaxy (Statler 1991). Each is likely to be correct in some circumstances. IC 1459's highly non-Gaussian LOSVDs originally motivated the second possibility, though there are compelling arguments in that case for the third; NGC 5982 is a good candidate for the fourth (Osterloo *et al.* 1994). For NGC 1700 the first option is the most likely explanation. Orbital projection effects are probably not at work here, first because of the evidence for near axisymmetry discussed above, and second because counterrotating cores produced by this effect almost never rotate about their apparent minor axes. A central disk is also probably not the culprit, given the low amplitude of the counterrotation and the modest $h_3$ values. The light profile is centrally cusped at the subarcsescond scale (Franx *et al.* 1989*a*), so unless the galaxy is dominated in its center by a dark halo with a large core radius, the circular velocity should be of the same order as the central dispersion, $\sim 200\,\mathrm{km\,s^{-1}}$. An admixture of a cold component with such a mean velocity to a slowly rotating hot population would produce highly skewed LOSVDs, which we do not observe. The essential distinction between NGC 1700 and IC 1459 is that the counterrotation in the latter is nearly an order of magnitude higher, $\sim 150\,\mathrm{km\,s^{-1}}$, much more typical of a cold disk, and the LOSVD asymmetry, as measured by the $h_3$ parameter, larger by more than a factor of two. For this same reason a major merger of spirals, which, as Schweizer (1990) has argued, would be prone to produce a strongly counterrotating core akin to IC 1459, is not a likely explanation. A disk could, in principle, still be dynamically viable if inclination were diluting the *projected* circular velocity by a factor of a few, but that would in turn imply an unprojected mean rotation at $2\,r_e$ of $\sim 500\,\mathrm{km\,s^{-1}}$, unless the galaxy were strongly twisted inside this radius. Moreover, if the galaxy is oblate, then at an observed $\epsilon = 0.3$ no inclination of the line of sight to the intrinsic short axis $< 50°$ is compatible with the known absence of ellipticals flatter than E6; this would imply that the maximum velocity projection effect is only a factor of 1.3. This latter constraint is not ironclad, however, and loosens considerably if the galaxy is triaxial. Finally, Bender & Surma (1992) call attention to the elevated $\mathrm{Mg_2}$ indices seen in the counterrotating cores of four ellipticals as a possible indication of major mergers; however, our preliminary analysis shows *no* elevation of $\mathrm{Mg_2}$ in the core of NGC 1700. There is thus no indication that the counterrotating material formed from the settled remains of a gas-rich spiral.

Balcells & Quinn (1990) simulate the accretion of a low-mass stellar system onto an oblate rotating primary, and find that counterrotating cores can be produced if the initial secondary orbit is retrograde with respect to the primary rotation. An important aspect of this process, further emphasized by Balcells (1991), is that much of the counterrotation in the final remnant owes to the stars of the *primary* that have had to absorb the orbital angular momentum of the secondary. Contrary to intuition, the core of the secondary is not tidally spun up during orbital decay; rather, it is spun *down* by a tendency to shed stars with high angular momentum. Thus the secondary core settles, barely rotating, into the center of the remnant, with much of the dynamical signature left in the stars of the primary, which acquire a skewed velocity distribution in the retrograde direction. Balcells (1991) obtains coefficients of skewness $\sim 0.2$, corresponding to $h_3 \sim 0.03$, very near the values we measure in the core of NGC 1700. Furthermore, Balcells & Quinn (1990) confirm Kormendy's (1984) conjecture that the central velocity dispersion of the remnant should be slightly depressed, because the central light comes to be dominated by the old core of a less massive system. We also observe a depression in the central dispersion of NGC 1700 (fig. 7). Note that this feature is not expected if counterrotation is due to a central disk, and is not observed in IC 1459. Thus the evidence very much favors the ingestion of a small stellar companion on a retrograde orbit as the cause of NGC 1700's counterrotating core.

### 4.3. Large Radii

At $R \approx 40''$ the photometric and kinematic structure of the galaxy change considerably. The isophotes become boxy, and both the photometric and kinematic major axes begin to rotate counterclockwise. The mean isophotal twist, compared to the inner regions, reaches 6° in our next-to-outermost and 15° in our outermost radial bins; the kinematic twist is roughly twice this. Finally, the outermost bin shows the velocity reversal on PA 225.

The dynamically simplest interpretation of the apparent twisting for $R \lesssim 55''$ is a radial increase of triaxiality. The data are consistent with a galaxy whose isodensity surfaces are aligned throughout, if the middle-to-long axis ratio $b/a$ begins to decrease outward at $R = 40''$ and the line of sight is not in one of the principal planes. The expected kinematic twist will naturally be in the right direction if the intrinsic short axis remains the axis of rotation. This model has the virtue of being a possible equilibrium configuration, since it requires neither intrinsic twists nor steep orbit population gradients. The restriction on the line of sight also somewhat favors an intrinsically flatter galaxy. On the other hand, if the main body of NGC 1700 is oblate, and intrinsically as flat as it appears, then the photometric twist can be explained only by a true warp of the equatorial plane. Triaxiality may not be required, but of course this is not an equilibrium configuration, unless elliptical galaxies have discrete warping modes as disks seem to have (*e.g.,* Sparke & Casertano 1988).

This discussion may be moot, though, since the isophote twist continues beyond $R \approx 60''$, where it seems certain that the galaxy is not in dynamical equilibrium. No equilibrium state would allow a reversal of the mean velocity to the northeast along PA 225 without a corresponding reversal to the southwest. The PA 225 profile resembles the U-shaped rotation curves seen in a number of disturbed systems (Borne & Hoessel 1988, Borne *et al.* 1994). These U-shaped profiles are a strong sign of dynamical interactions, but are extremely transient since they are produced by a dynamical friction "wake" and thus appear only in ongoing or very recent mergers. Alternatively, one is tempted to attribute the reversal to a long-lived tidal tail, but there is no photometric sign of any such feature on the northeast side of the galaxy in deep CCD frames (Schweizer 1990; 1995, private communication). The outer isophotes may suggest two broad tidal tails to the southeast and northwest (Schweizer 1990), but our outermost data points on PA 315 show that this material is rapidly rotating, contrary to what one would expect from tidal streamers drawn out mostly radially. The rapid rotation is more suggestive of a more rotationally-supported outer structure viewed nearly perpendicular to its rotation axis; Franx *et al.* (1989*a*) describe this as a "ring", although the radial density profile is not really known. A rapidly rotating outer structure could have been acquired through merger or ingestion of a smaller system, either by torquing up the outer part of the primary (Balcells & Quinn 1990) or by direct deposition of stars stripped from the secondary (*e.g.,* Statler 1988).

Whatever its origin, if the outer component is not aligned with the principal axes of the potential, then it will gradually become aligned due to differential precession and phase mixing. We can exploit the very regular structure of the galaxy interior to $R = 35''$ to place a lower limit on the time since the last major dynamical event. That the isophotes are well aligned and the velocity field is self-similar within this radius means that all transient structures inside $r = 6.6h^{-1}$ kpc have had time to phase-mix away.

In the epicyclic approximation, the precession frequency of a massless inclined ring in an oblate potential $\Phi(R, z)$ is

$$\omega_p = \left(\frac{\partial^2\Phi}{\partial z^2}\right)^{1/2} - \left(\frac{1}{R}\frac{\partial\Phi}{\partial R}\right)^{1/2}, \tag{10}$$

where the partial derivatives are evaluated on the circular orbit of radius $R$ at $z = 0$. Barring significant tangential anisotropy, we can assume that the potential is not far from logarithmic:

$$\Phi(R, z) = \frac{V_c^2}{2}\ln\left(R^2 + \frac{z^2}{q^2}\right), \tag{11}$$

implying that the precession time $T_p \equiv 2\pi/\omega_p$ is

$$T_p = \frac{\sqrt{2}\,\pi R}{\sigma}\left(\frac{1}{q} - 1\right)^{-1}, \tag{12}$$

where we have used the asymptotic relation $V_c = \sqrt{2}\,\sigma$ for the isothermal sphere. Since the precession time increases roughly linearly with $r$, a tilted structure will be phase-wrapped at a scale one-fourth its radius after $4T_p \equiv T_{\rm mix}$; we estimate this as the minimum time needed for such a structure to become indiscernible in projection. The flattening of the equipotentials $q$ is related to the flattening of the density surfaces $c/a \equiv q_\rho$ by (Binney & Tremaine 1986)

$$q_\rho^2 = q^2(2q^2 - 1). \tag{13}$$

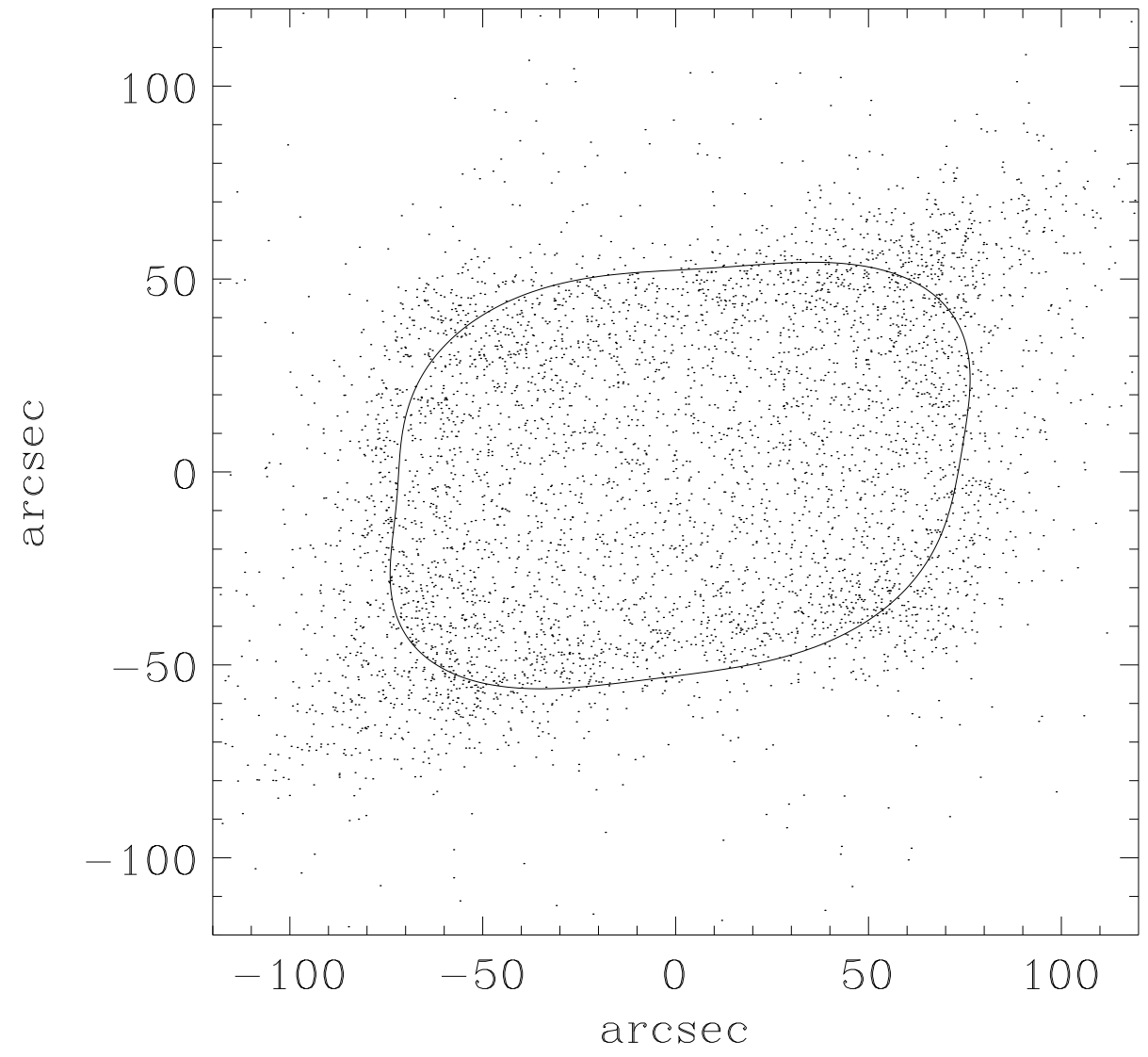

Fig. 15.— An inclined ring of massless test particles has been allowed to differentially precess in the $q = 0.9$ logarithmic potential for 1.4 mean precession periods. The configuration is viewed from within the equatorial plane of the potential, which is horizontal in the figure. The 22.89 $R$ mag/□″ isophote, as fitted by Franx *et al.* (1989*a*) is overplotted for comparison; the major axis of the inner isophotes is also horizontal.

Thus if mass and light are similarly flattened and the line of sight is in the equatorial plane, $q_\rho = 0.7$, $q = 0.9$, and $T_{\rm mix} = 5.3h^{-1}$ Gyr. The lower limit to the phase mixing time, however, is obtained if the galaxy is maximally flat ($q_\rho = 0.4$, $q = 0.73$), in which case $T_{\rm mix} = 2.2h^{-1}$ Gyr.

A simple numerical simulation helps to verify the approximate estimate of the mixing time. We follow the evolution of an inclined ring of 5000 non-interacting test particles in the oblate logarithmic potential (11). The particles are given an isotropic random velocity distribution, with dispersion $0.2V_c$, about the circular orbit of unit radius in the spherical ($q = 1$) potential. The ring is allowed to phase-mix in the spherical potential for 17.5 circular-orbit periods, then deposited into the oblate potential inclined to the equatorial plane. The inclination and velocity dispersion are chosen so that, edge-on, the particle ring resembles the photometric "ring" at $R \sim 80''$. For the $q = 0.9$ potential the inclination is 30° or 40°; for the $q = 0.73$ potential it is 45° so that, edge-on, it will appear inclined 30° to the major axis on a line of sight that gives the galaxy the correct apparent ellipticity. While this particle distribution is not representative of the distribution function in the galaxy, it allows us to follow differential precession locally; the scale-free nature of the potential then permits approximate scaling of the results to arbitrary radius since all frequencies scale as $r$. The simulation will tend to slightly overestimate the true mixing time since very eccentric orbits are neglected.

For these large inclinations, the average precession periods exceed the epicyclic calculation by about 10%. By visual inspection we find that the ring in either potential is essentially indistinguishable at any azimuth after $\sim 3.5$ true precession times, which in essence confirms the estimates above. Of particular interest is the strong resemblance of the incompletely mixed configurations, an example of which is shown in figure 15, to the outer isophotes of NGC 1700. Note that both the parallelogram-shaped isophotes and the outer "ring" are evident. The correspondence may be only circumstantial — and of course this simple model still does not explain the velocity reversal on PA 225 — but nevertheless suggests the likelihood that the outer asymmetrically boxy isophotes and the extensions to the northwest and southeast comprise a single dynamical structure, a differentially precessing disk or ring.

NGC 1700's last major merger must therefore have been quite a long time ago, otherwise it would not appear so photometrically and kinematically regular out to more than a dozen kpc. At the same time, it must not have been *too* far in the past, or we would no longer see the outer isophotal distortions. Not more than $1.5T_p$ should have elapsed for the material at $R \approx 70'' = 13h^{-1}$ kpc. Assuming the potential is roughly logarithmic (so that precession times scale as $r$), we estimate the time since the last major merger to be in the range $2.2 - 3.9h^{-1}$ Gyr. This is consistent with the $5.5 - 8.3$ Gyr estimate (using $h = 0.5$) of Schweizer & Seitzer (1992), which was obtained through similar considerations but without the benefit of stellar kinematics.

### 4.4. Origin in a Group Merger?

The evidence is strong that the boxy ring and the counterrotating core each formed through a merger event; but it cannot have been the *same* event. If, as the core LOSVDs suggest, counterrotation reflects the orbital angular momentum of a cannibalized companion, much of this angular momentum would have had to have been lost at large radii, and consequently deposited into the stars in the outer part of the primary. Balcells & Quinn (1990) find in their simulations of retrograde mergers that the primary is significantly spun down at all radii; yet we find that the rotation profile in NGC 1700 peaks outside $\sim 3\,r_e$, and the boxy ring — the other apparent signature of merging — seems to be the most rapidly prograde rotating part of the galaxy. A single merger of two systems would not deposit prograde-orbiting stars at large $r$ and retrograde-orbiting stars at small $r$. NGC 1700 must have merged with or ingested at least two other stellar systems to account for its present dynamical structure.

The individual mergers may or may not have occurred contemporaneously; we presently have no way to date the individual events, though stellar population studies may permit this in the future. It is inviting to speculate, however, that NGC 1700 may owe its present form to a compact group of galaxies that merged essentially at the same time. A systematic numerical study of group mergers has recently been undertaken by Weil (1995; see also Weil & Hernquist 1994). She finds three distinctive characteristics of the products of multiple mergers, as compared to pair mergers: (1) the remnants are predominantly oblate-triaxial; (2) their total angular momenta are closely aligned with their intrinsic short axes; and (3) their mean rotation peaks near or outside their effective radii. Our data strengthen the dynamical evidence for near-oblateness in NGC 1700, and the utter lack of minor-axis rotation indicates (though, admittedly, does not require) a small intrinsic kinematic misalignment (Franx *et al.* 1991). Also, the peak of the major-axis rotation curve occurs around $40'' = 2.9\,r_e$; in fact the maximum measured rotation occurs at our last measured point ($66'' = 4.7\,r_e$) on PA 315.[8] Superficially at least, NGC 1700 bears a close similarity to multiple merger remnants.

---

[8]To some extent, this only emphasizes the inadequacy of apparently global quantities like $v/\sigma$ that are actually determined

Clearly, more work on the theoretical side will be necessary to determine how multiple merger products might be unambiguously identified. We must emphasize that, however clear the evidence in NGC 1700 that past mergers have occurred, it still does not constitute proof that all, most, or in fact even any elliptical galaxies *formed* by multiple mergers. If NGC 1700's present morphology was determined by a group merger some $3h^{-1}$ Gyr ago, there is still no guarantee that the group might not have been dominated by a giant elliptical in the first place.

## 5. Conclusions

The evolutionary history of elliptical galaxies can perhaps be characterized as billions of years of collisionless boredom punctuated by moments of dissipative terror. Traumatic events — including, presumably, birth — leave their mark on the stellar dynamics of the system. We have demonstrated that the kinematic fields of ellipticals can be measured, with total errors $\lesssim 10\,\mathrm{km\,s^{-1}}$, well beyond the effective radius, potentially allowing new insights into the stressful lives of these apparently serene objects.

NGC 1700 is a galaxy whose morphology has been at least modified, and perhaps created, by mergers. There is clear evidence for past merging, both in the low velocity dispersion and asymmetric LOSVD of the counterrotating core, and in the apparent rapid rotation of the "boxy ring". At the same time, we have found that the main body of the galaxy, in the region $3'' \leq R \leq 35''$ ($0.6$ – $6.6h^{-1}$ kpc), appears kinematically well mixed, with a highly symmetric and radially self-similar mean velocity field. This indicates that the galaxy has had time since the last major merger to relax to a configuration that is probably nearly oblate, with rotation solely about the intrinsic short axis. This time must be longer than $2.2h^{-1}$ Gyr to allow for differential precession and phase mixing, but not longer than about $4h^{-1}$ Gyr, otherwise the outer photometric features would also have wound up and disappeared.

The same merger event cannot account for *both* the boxy ring and the counterrotating core; thus the present form of NGC 1700 must have resulted from a merger of at least three stellar systems. While we cannot establish whether these mergers occurred sequentially or simultaneously, simulations of group mergers produce objects strikingly similar to NGC 1700, both in intrinsic shape and kinematic structure. It is intriguing to speculate, though certainly not established, that NGC 1700 is one of a sub-family of nearly oblate, dynamically simple elliptical galaxies that formed by mergers of compact groups.

We thank the staff of the MMTO, especially Fred Chaffee, Craig Foltz, and Janet Robertson, for the allocation of time to this project and their patient assistance before, during, and after our run. We are also grateful to Jim Rose for lots of good advice. This work was supported in part by NASA Astrophysics Theory Program grant NAG 5-2860 (to TSS). TSH acknoweleges support from an NSF-NATO Post Doctoral Fellowship. We have made use of the NASA/IPAC Extragalactic Database (NED), which is operated by the Jet Propulsion Laboratory, Caltech, under contract with the National Aeronautics and Space Administration.

---

from local quantities to describe the structure of elliptical galaxies; in this case the amount of rotational support clearly varies with radius.